\documentclass[a4paper,11pt]{article}
\usepackage[utf8]{inputenc}
\usepackage{amsmath}
\usepackage{mathtools}
\usepackage{amsfonts}
\usepackage{amssymb,amsthm}
\usepackage{graphicx}
\graphicspath{{Figures/}}
\usepackage{epstopdf}
\usepackage{caption}
\usepackage{subcaption}
\usepackage{lscape}
\usepackage{xcolor}
\usepackage{multirow}
\usepackage{booktabs}
\usepackage{cellspace}
\newtheorem{prop}{Proposition}

\theoremstyle{remark}
\newtheorem{remark}{Remark}
\usepackage[font=small,labelfont=bf]{caption}

\newcommand{\mb}[1]{\mathbf{#1}}
\newcommand{\bs}[1]{\boldsymbol{#1}}
\newcommand{\Exp}{\mathbb{E}}
\newcommand{\Var}{\text{Var}}

\numberwithin{equation}{section}
\usepackage[toc]{appendix}

\DeclareMathOperator*{\argmax}{argmax}

\usepackage[a4paper, total={6.5in, 9.5in}]{geometry}

\usepackage{algorithm}
\usepackage{algpseudocode}

\usepackage{authblk}
\usepackage[authoryear]{natbib}
\usepackage{hyperref}
\definecolor{linkblue}{rgb}{0.05,0.20,0.50}
\hypersetup{
  colorlinks = true,
  linkcolor  = linkblue,   
  citecolor  = linkblue,   
  urlcolor   = {[rgb]{0.40,0.10,0.10}},
  linktoc    = all
}

\author{Giuseppe Buccheri\footnote{Department of Economics, University of Verona, giuseppe.buccheri@univr.it.}, Giacomo Bormetti\footnote{Department of Economics and Management, University of Pavia.},  Fulvio Corsi\footnote{Department of Economics, University of Pisa (corresponding author).} and Fabrizio Lillo\footnote{Department of Mathematics, Scuola Normale Superiore, Italy.}}

\renewcommand{\baselinestretch}{1.5}

\begin{document}

\title{Filtering and Smoothing with Score-Driven Models\footnote{We acknowledge financial support from the Italian Ministry MUR under the PRIN project ``Dynamic models for a fast changing world:\ An observation-driven approach to time varying parameters" (grant agreement n.\ 20205J2WZ4).\ FL acknowledges partial support by the European Program scheme ``INFRAIA-01-2018-2019: Research and Innovation action", grant agreement \#871042 'SoBigData++: European Integrated Infrastructure for Social Mining and Big Data Analytics.}
}

\date{\today}
\maketitle

\begin{abstract}
\renewcommand{\baselinestretch}{1}\normalsize
 Score-driven models are, by construction, purely predictive filters. When such models are read as filters for an underlying latent process, rather than as data generating processes, this is a restriction rather than a feature, because the contemporaneous and the future observations also carry information on the current state.
 Starting from the observation that the Kalman filter and smoother recursions for linear Gaussian models can be written in terms of the score of the conditional log-likelihood, and that the predictive step then takes the form of a score-driven recursion, we generalize score-driven models along this direction by deriving an update filter and a smoother, which exploit the contemporaneous observation and the whole sample respectively, together with the corresponding conditional variances.
 In extensive Monte Carlo analyses the update filter lowers the mean square error of the predictive filter by $3.5\%$ to $8.8\%$, and the smoother by $32\%$ to $44\%$, with the ordering holding in every replication; confidence bands built on the associated conditional variances attain their nominal coverage, whereas bands that ignore filtering uncertainty capture less than a third of it.
 Empirically, we demonstrate the benefits of employing score-driven models as filters rather than as purely predictive processes, showing that the resulting smoothed estimates align more closely with realized quantities than their predictive counterparts.
 
\vspace{0.6cm}

\end{abstract}

\newpage

\section{Introduction} 

GARCH-type models (\citealt{Engle}, \citealt{Bollerslev}) and, more generally, observation-driven models (\citealt{Cox}), are a class of dynamic econometric models where the time-varying parameters depend on past observations. When these models are regarded as data generating processes, the time-varying parameters are completely revealed by past observations, which encode all the relevant information. As such, there is no room for smoothing and the only form of uncertainty is that coming from the finite sample distribution of the maximum likelihood estimates, called parameter uncertainty. On the other hand, observation-driven models can be regarded as predictive filters, since the time-varying parameters are one-step-ahead predictable. In this case, conditionally on past observations, the state variables have a non-degenerate density, called filtering uncertainty. This means that exploiting information from contemporaneous and future observations would provide better estimates compared to one-step-ahead predictions. This idea was largely exploited by Daniel B.~Nelson, who explored the properties of the conditional covariances of misspecified GARCH filters under the assumption that data are generated by a continuous-time diffusion\footnote{The interpretation of GARCH processes as predictive filters is well described in this statement by \cite{NELSON199261}: \textit{``Note that our use of the term ‘estimate’ corresponds to its use in the filtering literature rather than the statistics literature; that is, an ARCH model with (given) fixed parameters produces ‘estimates’ of the true underlying conditional covariance matrix at each point in time in the same sense that a Kalman filter produces ‘estimates’ of unobserved state variables in a linear system''.}}; see \cite{NELSON199261}, \cite{NelsonFoster}, \cite{NELSON1995303} and \cite{NelsonSmooth}. In particular, \cite{NelsonSmooth} showed how to efficiently use information in both lagged and led GARCH residuals to estimate the spot volatility of a Brownian diffusion. 
Despite the considerable amount of observation-driven models proposed in the econometric literature, little attention has been paid to examining their properties as misspecified filters and, as a consequence, to using them for smoothing and assessing filtering uncertainty.\footnote{An important recent exception is \cite{BeutnerLinLucas}, whose results we discuss below.}

We aim to fill this gap by introducing a filtering and smoothing methodology for the class of score-driven models of \cite{GAS1} and \cite{Harvey_2013}. Score-driven models have been successfully applied in several areas of the recent econometric literature; see for instance \cite{GAS2}, \cite{HarveyLuati}, \cite{OhPatton}, \cite{BABII201947}, \cite{HafnerHerwartz}, \cite{CrealMoment}, \cite{Artemova2025} and \cite{BuccheriInvariance}.\footnote{An up-to-date bibliography of the literature on score-driven models is maintained at \url{http://www.gasmodel.com}.}
We start from the observation that the well-known Kalman filter and smoother recursions for linear Gaussian models can be written in terms of the score and a measure of curvature of the conditional log-likelihood function. In particular, the predictive filter has the form of a standard score-driven recursion, with a score normalization given by the conditional covariance of the underlying state variables. 
Since the predictive step of the Kalman filter, once written in terms of the score, is a score-driven recursion, the remaining Kalman recursions can be transported to the score-driven world by the same route. We replace the Gaussian score with the score of the observation density at hand, and obtain recursions that apply to a general nonlinear non-Gaussian model.
The formal relation between the Kalman predictive filter and score-driven models allows us, on the one hand, to provide a further justification for the updating rule adopted in score-driven models, and, on the other hand, to derive the score-driven counterparts of the remaining Kalman recursions, namely (i) an update filter, which exploits the contemporaneous observation; (ii) a smoother, which exploits the whole sample; and (iii) the conditional variances attached to the three estimates, from which confidence bands accounting for both parameter and filtering uncertainty can be constructed.
The resulting gain is far from marginal. In our simulations the update filter lowers the mean square error of the predictive filter by $3.5\%$ to $8.8\%$, and the smoother lowers it by $32\%$ to $44\%$, so that the information contained in the future of the sample is worth roughly six times the information contained in the present.

The problem of filtering and smoothing for general state-space models reduces to a multi-dimensional integral over the space spanned by the state-variables. Computationally intensive simulation techniques are typically employed to compute such integrals (see, e.g., \citealt{DurbinKoopman}). 
The main advantage of the proposed methodology is that its computational complexity is the same as in linear Gaussian models. The predictive and update filters are indeed computed iteratively through a simple forward recursion. Smoothing only requires an additional backward recursion, which updates the filtered estimates using the information of all the available data. 
In our experiments the backward pass costs less than two per cent of the forward one, since it evaluates no density and merely recycles the scores and the information quantities already computed going forward. Access to the whole sample, in other words, is obtained essentially for free once the filter has been run.

As evidenced above, correctly specified observation-driven models are only affected by parameter uncertainty, which comes from replacing the true static parameters with their maximum likelihood estimates. However, if observation-driven models are employed as misspecified filters, the latent state variables are not completely revealed by past observations. Thus, also filtering uncertainty is relevant when building confidence bands. While parameter uncertainty can be quantified through the methods developed, e.g., by \cite{PASCUAL20062293} and \cite{BLASQUES2016875}, it is less clear how one can take into account filtering uncertainty in observation-driven models. 
We propose a methodology to approximate in-sample and out-of-sample confidence bands, in a similar fashion to linear Gaussian models. Furthermore, depending on the needs of the user, the constructed confidence bands can account for filtering uncertainty only, for parameter uncertainty only, or for both parameter and filtering uncertainty.

Besides the work of Nelson mentioned above, observation-driven models have been employed as misspecified filters in other works. 
\cite{Blasques} proved that score-driven filters are locally optimal based on information theoretic criteria, while \cite{GAS3} showed through an extensive Monte Carlo study that misspecified score-driven models provide similar forecasting performances as correctly specified parameter-driven models. \cite{BeutnerLinLucas} study the in-fill asymptotics of score-driven models under general forms of mis-specification, showing that score-driven filters are consistent for the Kullback-Leibler optimal time-varying parameter path, obtaining the limiting distribution of the filtering errors, and deriving the observation-driven filter that minimizes the asymptotic filter error variance, which turns out to be score-driven when the predictive conditional density is correctly specified. Their results generalize Nelson's continuous-time findings and concern the predictive filter; they are in this sense complementary to the question we ask here, namely how much is gained by conditioning on the contemporaneous observation and on the whole sample rather than on the past alone.

\cite{Harvey_2013} proposed a related smoothing technique for a dynamic Student-$t$ location model by replacing the prediction error in the Kalman smoother recursions with the score of the Student-$t$ density. An application of such a technique can be found in \cite{Caivano2016}. Our approach is based on a representation of the Kalman filter and smoother recursions in terms of the score of the conditional log-likelihood which leads to different smoothing recursions. The latter are readily applicable to a general nonlinear non-Gaussian density. Our recursions are also related to earlier approximate filters for non-Gaussian state-space models, and in particular to the popular algorithm of \cite{Masreliez}, which likewise updates the state estimate through the score of the observation density.\footnote{\cite{Masreliez} derives his recursions under the assumption that the predictive density of the latent state, $p(\bs{\alpha}_t|\mb{Y}_{t-1})$, is Gaussian. Under such assumption the update is driven by the score of the marginal density $\int p(\mb{y}_t|\bs{\alpha}_t)p(\bs{\alpha}_t|\mb{Y}_{t-1})d\bs{\alpha}_t$, which is generally not available in closed form and must be evaluated by numerical convolution at each time step. His algorithm is moreover derived for location models. In the recursions proposed here, the score is that of the conditional density $p(\mb{y}_t|\mb{Y}_{t-1})$ evaluated at the predictive filter. As such, it is available analytically for all the models we consider, it requires no numerical integration, and it applies to location, scale and count data models alike.}

A closely related concern motivates \cite{ArtemovaESD}, who propose an extended score-driven dynamic factor model whose state equation includes the contemporaneous score in addition to the lagged one. Their starting point coincides with ours, namely that the factor of a standard score-driven model is predetermined and thus carries no filtering uncertainty. Their solution, however, is to modify the data generating process, obtaining a new class of models in which prediction and updating genuinely differ, whereas we leave the score-driven model untouched and read it as a misspecified filter. Their analysis is moreover confined to the factor setting and does not consider smoothing. \cite{LangeImplicit} likewise distinguish a prediction from an updating step in their class of implicit score-driven filters.

We assess the three estimates in an extensive Monte Carlo study covering location, scale and duration models with non-Gaussian observation densities. Beyond the average gains reported above, the ordering of the three estimates is remarkably stable. The mean square error of the smoother is below that of the update filter, which in turn is below that of the predictive filter, in \emph{every one} of the thousand replications and for every model considered. The size of the gain varies systematically with how informative a single observation is about the state, and grows with it, since the predictive filter is the only estimate that cannot exploit the contemporaneous observation. For the same set of models we construct confidence bands around the three estimates and find average coverage rates close to the nominal levels, the discrepancy never exceeding four percentage points and staying below one point for two of the four models. Bands that account for parameter uncertainty alone, by contrast, capture less than a third of the uncertainty they are meant to represent. Parameter uncertainty vanishes as the estimation sample grows, whereas filtering uncertainty does not vanish at all.

Our empirical illustration is devoted to showing the advantages of employing score-driven models as misspecified filters rather than purely predictive processes. We estimate the conditional covariance of daily open-to-close returns on Standard \& Poor's 500 constituents through the multivariate $t$-GAS model of \cite{GAS2}, and use the realized covariance computed from intraday transaction data as a proxy for the latent target, so that the three estimates can be compared through standard loss functions. The smoothed estimates are the only ones entering the model confidence set, in every portfolio and for both loss functions, and their advantage is stable as the cross-sectional dimension grows from five to twenty assets. This last point matters in practice, since our recursions remain computationally simple in high dimensions, whereas simulation-based alternatives become problematic; we quantify the two costs in Appendix \ref{app:cost}.

The rest of the paper is organized as follows: Section \ref{sec:ModelConstruction} introduces the methodology and provides the main theoretical results; Section \ref{sec:MonteCarlo} shows the results of the Monte Carlo study; in Section \ref{sec:empirics} the methodology is applied to a dataset of asset prices belonging to the S\&P 500 index; Section \ref{sec:Conclusions} concludes.

\section{Methodology}
\label{sec:ModelConstruction}

In this section, we discuss how, in the steady state, the classical Kalman filter and smoothing recursions for linear Gaussian models can be re-written in an alternative form that only involves the score of the conditional likelihood, the Fisher information matrix and a set of static parameters. Abstracting from the linear Gaussian setting, these recursions can be viewed as the approximate filtering and smoothing recursions for a non-Gaussian model by computing scores and information based on the non-Gaussian density. 

\subsection{Linear Gaussian models}
\label{sub:kf}

Our starting point is the well-known theory of linear Gaussian models. Let us consider the following state-space representation:
\begin{align}
\mb{y}_t &= \mb{Z}\bs{\alpha}_t+\bs{\epsilon}_t,\qquad \bs{\epsilon}_t \sim\text{NID}(\mb{0},\mb{H})\label{eq:lgmObs}\\
\bs{\alpha}_{t+1} &= \mb{c}+ \mb{B}\bs{\alpha}_t+\bs{\eta}_t,\qquad \bs{\eta}_t \sim\text{NID}(\mb{0},\mb{Q})\label{eq:lgmTrans}
\end{align} 
where $\bs{\alpha}_t\in\mathbb{R}^m$ is a vector of state variables and $\mb{y}_t\in\mathbb{R}^p$ is a vector of observations. The parameters $\mb{c}\in\mathbb{R}^{m}$, $\mb{Z}\in\mathbb{R}^{p\times m}$, $\mb{H}\in\mathbb{R}^{p\times p}$, $\mb{B}\in\mathbb{R}^{m\times m}$ and $\mb{Q}\in\mathbb{R}^{m\times m}$ are referred to as system matrices. Let $\mb{Y}_t$ denote the set of observations up to time $t$, namely $\mb{Y}_t=\{\mb{y}_1,\dots,\mb{y}_t\}$. We are interested in computing the mean and the variance of the underlying state variable $\bs{\alpha}_t$ based on the observations $\mb{y}_1,\dots,\mb{y}_t$ available at time $t$ (update step) and in computing the mean and the variance of $\bs{\alpha}_{t+1}$ based on the past observations $\mb{y}_1,\dots,\mb{y}_t$ (prediction step). We thus define:
\begin{align}
\mb{a}_{t|t}=\Exp [\bs{\alpha}_t|\mb{Y}_t],\qquad & \mb{P}_{t|t}=\Var [\bs{\alpha}_t|\mb{Y}_t]\\
\mb{a}_{t+1}=\Exp[\bs{\alpha}_{t+1}|\mb{Y}_t],\qquad & \mb{P}_{t+1}=\Var[\bs{\alpha}_{t+1}|\mb{Y}_t]
\end{align}  
The Kalman filter allows one to compute recursively $\mb{a}_{t|t}$, $\mb{P}_{t|t}$, $\mb{a}_{t+1}$ and $\mb{P}_{t+1}$. Assuming $\bs{\alpha}_1\sim\text{N}(\mb{a}_1,\mb{P}_1)$, where $\mb{a}_1$ and $\mb{P}_1$ are known, for $t=1,\dots,n$, we have (see, e.g., \citealt{Harvey}, \citealt{DurbinKoopman}):
\begin{align}
\mb{v}_t  = \mb{y}_t-\mb{Z}\mb{a}_t , \qquad \qquad & 
\mb{F}_t = \mb{Z}\mb{P}_t\mb{Z}'+\mb{H} \label{eq:kfStandard1} \\
\mb{a}_{t|t} = \mb{a}_t + \mb{P}_t\mb{Z}'\mb{F}_t^{-1}\mb{v}_t, \qquad \qquad & 
\mb{P}_{t|t} = \mb{P}_t - \mb{P}_t\mb{Z}'\mb{F}_t^{-1}\mb{Z}\mb{P}_t \label{eq:kfStandard2}\\
\mb{a}_{t+1} = \mb{c}+ \mb{B}\mb{a}_t + \mb{K}_t\mb{v}_t \label{eq:kfStandard3}, \qquad \qquad &
\mb{P}_{t+1} = \mb{B}\mb{P}_t(\mb{B}-\mb{K}_t\mb{Z})'+\mb{Q}
\end{align}
where $\mb{K}_t=\mb{B}\mb{P}_t\mb{Z}'\mb{F}_t^{-1}$. The conditional log-likelihood is normal and is given by:
\begin{equation}
\log p(\mb{y}_t|\mb{Y}_{t-1})=\text{const}-\frac{1}{2}\left(\log |\mb{F}_t| + \mb{v}_t'\mb{F}_t^{-1}\mb{v}_t\right)
\label{eq:ll} 
\end{equation}
The smoothed estimates $\hat{\bs{\alpha}}_t=\Exp[\bs{\alpha}_t|Y_n]$, $\hat{\mb{P}}_t=\Var[\bs{\alpha}_t|Y_n]$, $n>t$, can instead be computed through the following set of backward recursions:
\begin{align}
\mb{r}_{t-1}  = \mb{Z}'\mb{F}_t^{-1}\mb{v}_t + \mb{L}_t'\mb{r}_t, \qquad \qquad & 
\mb{N}_{t-1}  = \mb{Z}'\mb{F}_t^{-1}\mb{Z} + \mb{L}_t'\mb{N}_t \mb{L}_t \label{eq:ksStandard1}\\
\hat{\bs{\alpha}}_t  =\mb{a}_t + \mb{P}_t\mb{r}_{t-1}, \qquad \qquad &
\hat{\mb{P}}_t  = \mb{P}_t-\mb{P}_t\mb{N}_{t-1}\mb{P}_t \label{eq:ksStandard2}
\end{align}
where $\mb{L}_t=\mb{B}-\mb{K}_t\mb{Z}$, $\mb{r}_n=\mb{0}$, $\mb{N}_n=\mb{0}$ and $t=n,\dots,1$. All the conditional distributions involved are Gaussian: $\bs{\alpha}_{t+1}|\mb{Y}_t$ has mean and variance $(\mb{a}_{t+1}, \mb{P}_{t+1})$, whereas $\bs{\alpha}_{t}|\mb{Y}_t$ and $\bs{\alpha}_{t}|\mb{Y}_n$ have mean and variance $(\mb{a}_{t|t}, \mb{P}_{t|t})$ and $(\hat{\bs{\alpha}}_t, \hat{\mb{P}}_t)$, respectively.  

\subsection{Score-driven representation of the Kalman predictive filter}
\label{sub:genKf}
  
Let us denote by  $\bs{\nabla}_t = \frac{\partial\log p(\mb{y}_t|\mb{Y}_{t-1})}{\partial \mb{a}_t}$ the score of the conditional log-likelihood computed with respect to the predictive filter $\mb{a}_t$. From Eq. \eqref{eq:ll}, it readily follows that in linear Gaussian models the score $\bs{\nabla}_t$ is given by $ \bs{\nabla}_t = \mb{Z}'\mb{F}_t^{-1}\mb{v}_t$.
Let us also denote by $\bs{\mathcal{I}}_{t}=\Exp[{\bs{\nabla}_t\bs{\nabla}_t'}|\mb{Y}_{t-1}]$ the Fisher information matrix, which may be time-varying. 
In linear Gaussian models we have\footnote{Note that in linear Gaussian models the Fisher information matrix coincides with the negative Hessian, thus we could express the recursive equations in terms of both quantities.}  $\bs{\mathcal{I}}_t = \mb{Z}'\mb{F}_t^{-1}\mb{Z}$.

Therefore, it is possible to re-write the general Kalman filter and smoother equations in terms of $\bs{\nabla}_t$ and $\bs{\mathcal{I}}_t$. Indeed, Eq. \eqref{eq:kfStandard2}, \eqref{eq:kfStandard3} become:
\begin{align}
	\mb{a}_{t|t} & = \mb{a}_t + \mb{P}_t\bs{\nabla}_t \label{eq:kfGeneral2}\\
	\mb{a}_{t+1} & = \mb{c}+ \mb{B}\mb{a}_t + \mb{B}\mb{P}_t\bs{\nabla}_t \label{eq:kfGeneral3}
\end{align}
whereas the equations for the conditional variance become:
\begin{align}
	\mb{P}_{t|t} &= \mb{P}_t - \mb{P}_t\bs{\mathcal{I}}_t\mb{P}_t \label{eq:pttGeneral}\\
	\mb{P}_{t+1} &= \mb{B}\mb{P}_t(\mb{B}-\mb{B}\mb{P}_t\bs{\mathcal{I}}_t)'+\mb{Q} \label{eq:pt1General}
\end{align}
Similarly, the backward smoothing recursions \eqref{eq:ksStandard1}-\eqref{eq:ksStandard2} can be written as:
\begin{align}
	\mb{r}_{t-1} &= \bs{\nabla}_t + (\mb{I}-\mb{P}_t\bs{\mathcal{I}}_t)'\mb{B}'\mb{r}_t \label{eq:ksGeneral1}\\
	\hat{\bs{\alpha}}_t & =\mb{a}_t + \mb{P}_t\mb{r}_{t-1} \label{eq:ksGeneral2}
\end{align}
and:
\begin{align}
	\mb{N}_{t-1} &= \bs{\mathcal{I}}_t + (\mb{I}-\mb{P}_t\bs{\mathcal{I}}_t)'\mb{B}'\mb{N}_t\mb{B}(\mb{I}-\mb{P}_t\bs{\mathcal{I}}_t) \label{eq:NtGeneral} \\
	\hat{\mb{P}}_t &= \mb{P}_t-\mb{P}_t\mb{N}_{t-1}\mb{P}_t \label{eq:hatPtGeneral}
\end{align}
with $\mb{r}_n=\mb{0}$, $\mb{N}_n=\mb{0}$ and $t=n,\dots,1$.

The recursions above still involve the time-varying matrix $\mb{P}_t$ and, through it, a time-varying gain. Specializing them to the steady state, in which $\mb{P}_t$ is replaced by its limit, yields recursions whose coefficients no longer depend on time. The following proposition gives the resulting form.

\begin{prop}
	In the \emph{steady state}, Eq. (\ref{eq:kfStandard2}), (\ref{eq:kfStandard3}), (\ref{eq:ksStandard1}), (\ref{eq:ksStandard2}) can be written as:
	\begin{align}
		\mb{a}_{t|t} &= \mb{a}_t+\mb{B}^{-1}\mb{A}\bs{\nabla}_t \label{eq:kfNew1} \\
		\mb{a}_{t+1} &= \mb{c} + \mb{B}\mb{a}_t+\mb{A}\bs{\nabla}_t \label{eq:kfNew2}
	\end{align}
	and
	\begin{align}
		\mb{r}_{t-1} &= \bs{\nabla}_t+(\mb{B}-\mb{A}\bs{\mathcal{I}})'\mb{r}_t \label{eq:ksNew1}\\
		\hat{\bs{\alpha}}_t &= \mb{a}_t+\mb{B}^{-1}\mb{A}\mb{r}_{t-1} \label{eq:ksNew2}
	\end{align}
	where 
	$\bs{\mathcal{I}} = \mb{Z}'\mb{\bar{F}^{-1}}\mb{Z}$, $\mb{\bar{F}}=\mb{Z}\mb{\bar{P}}\mb{Z}' + \mb{H}$, $\mb{A}= \mb{B}\mb{\bar{P}}$ and $\mb{\bar{P}}$ is the steady state variance matrix which is the solution of the matrix Riccati equation:
	\begin{equation}
		\mb{\bar{P}} = \mb{B}\mb{\bar{P}}\mb{B}' - \mb{B}\mb{\bar{P}}\mb{Z}'\mb{\bar{F}^{-1}}\mb{Z}\mb{\bar{P}}\mb{B}' + \mb{Q} \label{eq:riccati}
	\end{equation}
	\label{prop:steadyState}
\end{prop}
\noindent The proof is reported in Appendix \ref{app:prop}.

Note that a steady state solution exists whenever the system matrices are constant (\citealt{Harvey}, \citealt{DurbinKoopman}). In this case, the variance matrix $\mb{P}_t$ converges to $\mb{\bar{P}}$ at an exponential rate. Throughout, we assume that the transition matrix $\mb{B}$ is invertible, so that the update and smoothing recursions (\ref{eq:kfNew1}), (\ref{eq:ksNew2}) are well defined. The role played by the steady state is worth stressing. It is precisely the convergence of $\mb{P}_t$ to the constant matrix $\mb{\bar{P}}$ that makes the gain $\mb{A}=\mb{B}\mb{\bar{P}}$ time-invariant.
The  Kalman recursions for the mean are re-parameterized in terms of the score $\bs{\nabla}_t$ and the Fisher information matrix $\bs{\mathcal{I}}$. In the steady state, this representation is equivalent to the one in equations (\ref{eq:kfStandard2}), (\ref{eq:kfStandard3}) and (\ref{eq:ksStandard1}), (\ref{eq:ksStandard2}). However, it is more general, as it only relies on the predictive density $p(\mb{y}_t|\mb{Y}_{t-1})$. In principle, the forward recursions (\ref{eq:kfNew1}), (\ref{eq:kfNew2}) and the backward recursions (\ref{eq:ksNew1}), (\ref{eq:ksNew2}) can be applied to any state-space model for which a predictive density $p(\mb{y}_t|\mb{Y}_{t-1})$ is defined.

\subsection{Score-driven filtering and smoothing recursions}
\label{sub:sdsRec}

Since the gain is time-invariant, the predictive filter in the steady state (\ref{eq:kfNew2}) is an autoregressive recursion with static parameters, driven by the score of the conditional likelihood, i.e., it has the form of the score-driven models of \cite{GAS1} and \cite{Harvey_2013}. Thus, if one looks at score-driven models as filters, it turns out that the score-driven filter (\textsc{SDF} hereafter) is optimal in the case of linear Gaussian models. In the case of nonlinear non-Gaussian models, the SDF can be regarded as an approximate nonlinear filter. The main difference with respect to the Kalman filter is that the Gaussian score is replaced by the score of the conditional density, thus providing robustness to non-Gaussianity.

Based on the same principle, we introduce an approximate nonlinear update filter, which estimates the time-varying parameters using the observations up to and including the contemporaneous one, and an approximate nonlinear smoother, which uses all available observations. In linear Gaussian models, the Kalman update and smoothing recursions return the conditional mean of the state and are therefore optimal in mean square error among all estimators; when the Gaussianity assumption is dropped, they remain the minimum variance linear unbiased estimators (MVLUE) of the state. Thus, we define our update filter and smoother in such a way that they coincide with the latter in this specific case. In the case of nonlinear non-Gaussian models, they maintain the same simple form of the Kalman forward and backward recursions but replace the Gaussian score with the one of the non-Gaussian density.

Let us assume that observations $\mb{y}_t\in\mathbb{R}^p$, $t=1,\dots,n$, are generated by the following observation density: 
\begin{equation}
	\mb{y}_t|\mb{f}_t \sim p(\mb{y}_t|\mb{f}_t,\bs{\theta})
\end{equation}
where $\mb{f}_t\in\mathbb{R}^k$ is a vector of $\mb{Y}_{t-1}-$ measurable time-varying parameters and $\bs{\theta}$ is a vector of static parameters. We generalize the filtering and smoothing recursions (\ref{eq:kfNew1})-(\ref{eq:ksNew2}) for the measurement density $p(\mb{y}_t|\mb{f}_t,\bs{\theta})$ as:
\begin{align}
	\mb{f}_{t|t} &= \mb{f}_t + \mb{B}^{-1}\mb{A}\bs{\nabla}_t \label{eq:sdsU}\\
	\mb{f}_{t+1} &= \bs{\omega} + \mb{A}\bs{\nabla}_t + \mb{B}\mb{f}_t \label{eq:sdsF}
\end{align}
$t=1,\dots,n$ and:
\begin{align}
	\mb{r}_{t-1} &= \bs{\nabla}_t+(\mb{B}-\mb{A}\bs{\mathcal{I}}_{t})'\mb{r}_{t} \label{eq:sdsR}\\
	\mb{\hat{f}}_t &= \mb{f}_t + \mb{B}^{-1}\mb{A}\mb{r}_{t-1} \label{eq:sdsS}
\end{align}
where $\mb{r}_n=\mb{0}$ and $t=n,\dots,1$. The predictive filter in equation (\ref{eq:sdsF}) has the same form of score-driven models. The term $\bs{\nabla}_t$ is now the score of the measurement density $p(\mb{y}_t|\mb{f}_t,\bs{\theta})$, namely $\bs{\nabla}_t = \frac{\partial\log p(\mb{y}_t|\mb{f}_t,\bs{\theta})}{\partial \mb{f}_t}$,
and $\bs{\mathcal{I}}_{t}=\Exp[\bs{\nabla}_t\bs{\nabla}_t'|\mb{Y}_{t-1}]$ is the information matrix. Note that, while in the linear Gaussian steady state the information matrix is constant, in nonlinear non-Gaussian models it generally depends on $\mb{f}_t$ and is therefore time-varying; we retain the subscript $t$ to emphasize this.
The vector $\bs{\omega}\in\mathbb{R}^k$ and the two matrices $\mb{A},\mb{B}\in\mathbb{R}^{k\times k}$ are static parameters included in $\bs{\theta}$. They are estimated by maximizing the log-likelihood, namely:
\begin{equation}
	\tilde{\bs{\theta}} = \argmax\limits_{\bs{\theta}}\sum_{t=1}^n \log p(\mb{y}_t|\mb{f}_t,\bs{\theta})
\end{equation}
Thus, one can run the backward smoothing recursions (\ref{eq:sdsR}), (\ref{eq:sdsS}) after computing the forward filtering recursions (\ref{eq:sdsU}), (\ref{eq:sdsF}), in a similar fashion to Kalman filter and smoothing recursions. Note that the above recursions are nonlinear, as the score of a non-Gaussian density is typically nonlinear in the observations. The filter $\mb{f}_{t|t}$ in equation (\ref{eq:sdsU}) allows one to update the current estimate $\mb{f}_t$ once a new observation $\mb{y}_t$ becomes available. While going backward, the smoothing recursions (\ref{eq:sdsR}), (\ref{eq:sdsS}) update the two filters $\mb{f}_t$ and $\mb{f}_{t|t}$ using all available observations. Smoothed estimates $\mb{\hat{f}}_t$ are generally less noisy than filtered estimates $\mb{f}_{t|t}$, $\mb{f}_t$ and provide a more accurate reconstruction of the time-varying parameters. 

It is a standard practice in score-driven models replacing the score $\bs{\nabla}_t$ with the scaled score $\mb{s}_t=\mb{S}_t\bs{\nabla}_t$. The role of the scaling matrix $\mb{S}_t$ is to take into account the curvature of the log-likelihood function. \cite{GAS1} discussed several choices of $\mb{S}_t$ based on inverse powers of the information matrix $\bs{\mathcal{I}}_{t}$. For instance, given a normal density with time-varying variance, if $\mb{S}_t=\bs{\mathcal{I}}_{t}^{-1}$, one recovers the standard GARCH model. The filtering and smoothing recursions (\ref{eq:sdsU})-(\ref{eq:sdsS}) are obtained if one sets $\mb{S}_t$ equal to the identity matrix. When using a scaled score $\mb{s}_t$, the filtering recursions (\ref{eq:sdsU}), (\ref{eq:sdsF}) become:
\begin{align}
	\mb{f}_{t|t} &= \mb{f}_t + \mb{B}^{-1}\mb{A} \mb{s}_t \label{eq:sdsUgen}\\
	\mb{f}_{t+1} &= \bs{\omega} + \mb{A} \mb{s}_t + \mb{B}\mb{f}_t \label{eq:sdsFgen}
\end{align}
Since the score is now scaled by $\mb{S}_t$, the term $\mb{A}\bs{\mathcal{I}}_{t}$ in equation (\ref{eq:sdsR}) has to take into account the new normalization, and $\mb{A}$ is replaced by $\mb{A}\mb{S}_t$, as shown in Remark \ref{rem:scaling} below. We thus obtain the general backward smoothing recursions:
\begin{align}
	\mb{r}_{t-1} &= \mb{s}_t+(\mb{B}-\mb{A}\mb{S}_t\bs{\mathcal{I}}_{t})'\mb{r}_{t} \label{eq:sdsR_gen}\\ 
	\mb{\hat{f}}_t &= \mb{f}_t + \mb{B}^{-1}\mb{A}\mb{r}_{t-1} \label{eq:sdsS_gen}
\end{align}
For instance, if $\mb{S}_t=\bs{\mathcal{I}}_{t}^{-1}$, we obtain:
\begin{align}
	\mb{r}_{t-1} &= \mb{s}_t+(\mb{B}-\mb{A})'\mb{r}_{t} \\
	\mb{\hat{f}}_t &= \mb{f}_t + \mb{B}^{-1}\mb{A}\mb{r}_{t-1} 
\end{align}
that is, the information matrix $\bs{\mathcal{I}}_{t}$ disappears because its effect is already taken into account when scaling the score. If $\mb{S}_t=\bs{\mathcal{I}}_{t}^{-1/2}$, we get:
\begin{align}
	\mb{r}_{t-1} &= \mb{s}_t+(\mb{B}-\mb{A}\bs{\mathcal{I}}_{t}^{1/2})'\mb{r}_{t} \\
	\mb{\hat{f}}_t &= \mb{f}_t + \mb{B}^{-1}\mb{A}\mb{r}_{t-1} 
\end{align}

\begin{remark}
	The form of the recursions (\ref{eq:sdsR_gen}), (\ref{eq:sdsS_gen}) is pinned down by two distinct requirements. Evaluating them at $t=n$ determines the driving term of equation (\ref{eq:sdsR_gen}) and the absence of any rescaling in equation (\ref{eq:sdsS_gen}). Since $\mb{r}_n=\mb{0}$, we obtain $\mb{r}_{n-1}=\mb{s}_n$ and hence $\mb{\hat{f}}_n=\mb{f}_n+\mb{B}^{-1}\mb{A}\mb{s}_n$, which coincides with the update filter $\mb{f}_{n|n}$ in equation (\ref{eq:sdsUgen}), as the smoother and the update filter do in the Kalman case. This leaves unconstrained the matrix $\mb{L}_t$ entering equation (\ref{eq:sdsR_gen}) through the term $\mb{L}_t'\mb{r}_t$, since that term vanishes at $t=n$. Its form is instead dictated by the role it plays in the backward recursion. In the linear Gaussian case of Section \ref{sub:kf}, $\mb{L}_t=\mb{B}-\mb{K}_t\mb{Z}$ is the derivative of the prediction $\mb{a}_{t+1}$ with respect to $\mb{a}_t$, and the same interpretation applies here. Differentiating equation (\ref{eq:sdsFgen}) and using $\Exp[\partial\bs{\nabla}_t/\partial\mb{f}_t'|\mb{Y}_{t-1}]=-\bs{\mathcal{I}}_{t}$ gives $\mb{L}_t=\mb{B}-\mb{A}\mb{S}_t\bs{\mathcal{I}}_{t}$, which reduces to the $\mb{B}-\mb{A}\bs{\mathcal{I}}_{t}$ of equation (\ref{eq:sdsR}) when $\mb{S}_t$ is the identity matrix.
	\label{rem:scaling}
\end{remark}

From a computational point of view, the backward recursions (\ref{eq:sdsR_gen}), (\ref{eq:sdsS_gen}) are simple since $\mb{s}_t$ and $\bs{\mathcal{I}}_{t}$ are typically available from the forward filtering recursion. We term the approximate smoother obtained through recursions (\ref{eq:sdsR_gen}), (\ref{eq:sdsS_gen}) as Score-Driven Smoother (SDS). In essence, for any score-driven model, one can devise a companion SDS recursion that only requires $\mb{s}_t$, $\bs{\mathcal{I}}_{t}$ and the static parameters, as estimated through the SDF. Note that the forward recursion (\ref{eq:sdsUgen}) is the analogue of recursion (\ref{eq:kfStandard2}) in the Kalman filter and allows one to update SDF estimates once a new observation $\mb{y}_t$ becomes available. We denote the approximate Score-Driven Update filter (\ref{eq:sdsUgen}) by SDU. The proposed methodology is summarized in Algorithm \ref{alg:sds}.

\begin{algorithm}[htbp]
\caption{Score-driven filtering and smoothing.}
\label{alg:sds}
\begin{algorithmic}[1]
	\State Estimate the static parameters,
	\begin{equation*}
		\tilde{\bs{\theta}} = \argmax\limits_{\bs{\theta}}\sum_{t=1}^n \log p(\mb{y}_t|\mb{f}_t,\bs{\theta})
	\end{equation*}
	\For{$t=1,\dots,n$} \Comment{forward pass}
		\State $\mb{f}_{t|t} = \mb{f}_t + \tilde{\mb{B}}^{-1}\tilde{\mb{A}}\mb{s}_t$ \Comment{update filter, SDU}
		\State $\mb{f}_{t+1} = \tilde{\bs{\omega}} + \tilde{\mb{A}}\mb{s}_t + \tilde{\mb{B}}\mb{f}_t$ \Comment{predictive filter, SDF}
	\EndFor
	\State $\mb{r}_n=\mb{0}$
	\For{$t=n,\dots,1$} \Comment{backward pass}
		\State $\mb{r}_{t-1} = \mb{s}_t+(\tilde{\mb{B}}-\tilde{\mb{A}}\mb{S}_t\bs{\mathcal{I}}_{t})'\mb{r}_{t}$
		\State $\mb{\hat{f}}_t = \mb{f}_t + \tilde{\mb{B}}^{-1}\tilde{\mb{A}}\mb{r}_{t-1}$ \Comment{smoother, SDS}
	\EndFor
\end{algorithmic}
\end{algorithm}

\begin{remark}
	When the information matrix $\bs{\mathcal{I}}_t$ is singular, which happens whenever the number of time-varying parameters exceeds the dimension of the signal entering the observation density, the matrix $\mb{A}$ is not uniquely identified.\footnote{We thank Andrew Harvey for pointing this out.} The filters (\ref{eq:sdsUgen}), (\ref{eq:sdsFgen}) are unaffected, since they depend on $\mb{A}$ only through the identified combination, and so is the backward recursion (\ref{eq:sdsR_gen}). The smoothed estimate (\ref{eq:sdsS_gen}), by contrast, depends on $\mb{A}$ itself, so that different admissible choices deliver different smoothed estimates. We provide more details on this point in Appendix \ref{app:ident}.
	\label{rem:ident}
\end{remark}

\subsection{Examples of SDS recursions}
\label{sec:Examples}

In this section we provide several examples of SDS estimates. As a first step, we focus on two volatility models that are quite popular in the econometric literature, namely the GARCH model of \cite{Bollerslev} and the Beta-$t$-GARCH model of \cite{Harvey_2013}. Both are score-driven models and can therefore be treated within our framework. As a third example, we present an AR(1) model with a score-driven autoregressive coefficient. The time-varying autoregressive coefficient allows one to capture temporal variations in persistence, as well as nonlinear dependencies (\citealt{BlasquesNonlinear}). Autoregressive models with time-varying coefficients have been employed by \cite{DELLEMONACHE2017482} and \cite{SHARK} for inflation and volatility forecasting, respectively. 

One of the advantages of the SDS recursions (\ref{eq:sdsR_gen}), (\ref{eq:sdsS_gen}) is that they maintain the same simple form when $\mb{f}_t\in\mathbb{R}^k$, $k>1$, is a vector containing multiple time-varying parameters. In this multivariate setting, the use of simulation-based techniques would be highly computationally demanding. In order to test the SDS in a multivariate setting, we consider the $t$-GAS model of \cite{GAS2}, a conditional correlation model for heavy-tail returns. In this model the number of time-varying parameters grows as the square of the number of assets, and it therefore provides an interesting multivariate framework in which to assess the performance of the SDS.  

\subsubsection{GARCH-SDS}
Consider the model:
\begin{equation}
	y_t = \sigma_t\epsilon_t, \quad \epsilon_t\sim\text{NID}(0,1)
\end{equation}
The predictive density is thus:
\begin{equation}
	p(y_t|\sigma_t^2) = \frac{1}{\sqrt{2\pi\sigma_t^2}}e^{-\frac{y_t^2}{2\sigma_t^2}}
\end{equation}
Setting $f_t=\sigma_t^2$ and $S_t=\bs{\mathcal{I}}_{t}^{-1}$, equations (\ref{eq:sdsUgen}), (\ref{eq:sdsFgen}) reduce to:
\begin{align}
	f_{t|t} &= f_t + B^{-1}A(y_t^2-f_t)\\
	f_{t+1} &= \omega + A(y_t^2-f_t) + Bf_t \label{eq:garch}
\end{align}
In particular, the predictive filter (\ref{eq:garch}) is the standard GARCH(1,1) model. The smoothing recursions (\ref{eq:sdsR_gen}), (\ref{eq:sdsS_gen}) reduce to:
\begin{align}
	r_{t-1} &=  y_t^2-f_t + (B-A)r_{t} \\
	\hat{f}_t &= f_t + B^{-1}Ar_{t-1} 
\end{align}
$t=n,\dots,1$.

\subsubsection{Beta-$t$-GARCH-SDS}
Consider the model:
\begin{equation}
	y_t = \sigma_t\epsilon_t, \quad \epsilon_t\sim t_{\nu}
\end{equation}
where $t_{\nu}$ denotes a standard Student $t$ distribution with $\nu$ degrees of freedom. The predictive density is thus:
\begin{equation}
	p(y_t|\sigma_t^2,\nu) = \frac{\Gamma\left(\frac{\nu+1}{2}\right)}{\Gamma\left(\frac{\nu}{2}\right)\sqrt{\pi\nu\sigma_t^2}}\left(1+\frac{y_t^2}{\nu\sigma_t^2}\right)^{-\frac{\nu+1}{2}}
\end{equation}
Setting $f_t=\sigma_t^2$ and $S_t=\bs{\mathcal{I}}_{t}^{-1}$, equations (\ref{eq:sdsUgen}), (\ref{eq:sdsFgen}) reduce to:
\begin{align}
	f_{t|t} &= f_t + B^{-1}A(w_ty_t^2-f_t)\\
	f_{t+1} &= \omega + A(w_ty_t^2-f_t) + Bf_t\label{eq:betat}
\end{align}
where the weights
\begin{equation}
	w_t = \frac{\nu+1}{\nu+y_t^2/f_t} \label{eq:betatWeights}
\end{equation}
follow from the score of the Student $t$ density, and the constant $(\nu+3)/\nu$ arising from the inverse information has been absorbed into $A$. In particular, the predictive filter (\ref{eq:betat}) is the Beta-$t$-GARCH model of \cite{Harvey_2013}. The weights (\ref{eq:betatWeights}) downweight large realizations of $y_t$, which is what makes the filter robust to outliers, and as $\nu\rightarrow\infty$ one has $w_t\rightarrow1$ and the Gaussian recursions of the previous example are recovered. The smoothing recursions (\ref{eq:sdsR_gen}), (\ref{eq:sdsS_gen}) reduce to:
\begin{align}
	r_{t-1} &=  w_ty_t^2-f_t + (B-A)r_{t} \\
	\hat{f}_t &= f_t + B^{-1}Ar_{t-1}
\end{align}
$t=n,\dots,1$.

\subsubsection{AR(1)-SDS}
Consider the model:
\begin{equation}
	y_t = \mu + \alpha_ty_{t-1} + \epsilon_t,\quad \epsilon_t\sim\text{N}(0,q^2)
\end{equation}
The predictive density is thus given by:
\begin{equation}
	p(y_t|\alpha_t)=\frac{1}{\sqrt{2\pi} q}\exp{\left[-\frac{1}{2}\left(\frac{y_t-\mu-\alpha_ty_{t-1}}{q}\right)^2\right]}
\end{equation}
Setting $f_t=\alpha_t$ and $S_t=\bs{\mathcal{I}}_{t}^{-1}$, equations (\ref{eq:sdsUgen}), (\ref{eq:sdsFgen}) reduce to:
\begin{align}
	f_{t|t} &= f_t + B^{-1}A\left(\frac{y_t-\mu-f_ty_{t-1}}{y_{t-1}}\right)\\
	f_{t+1} &= \omega  + A\left(\frac{y_t-\mu-f_ty_{t-1}}{y_{t-1}}\right) + Bf_t 
\end{align}
while the smoothing recursions (\ref{eq:sdsR_gen}), (\ref{eq:sdsS_gen}) reduce to:
\begin{align}
	r_{t-1} &=  \left(\frac{y_t-\mu-f_ty_{t-1}}{y_{t-1}}\right) + (B-A)r_{t} \\
	\hat{f}_t &= f_t + B^{-1}Ar_{t-1} 
\end{align}
$t=n,\dots,1$.

\subsubsection{$t$-GAS-SDS}
\label{sec:tGAS}
Let $\mb{y}_t\in\mathbb{R}^p$ denote a vector of demeaned daily log-returns. Consider the following observation density:
\begin{equation}
	p(\mb{y}_t|\mb{V}_t,\nu) = \frac{\Gamma((\nu+p)/2)}{\Gamma(\nu/2)[(\nu-2)\pi]^{p/2}|\mb{V}_t|^{1/2}}\left[1+\frac{\mb{y}_t'\mb{V}_t^{-1}\mb{y}_t}{(\nu-2)}\right]^{-\frac{\nu+p}{2}}
	\label{eq:obs_tGAS}
\end{equation}
where $\mb{V}_t\in\mathbb{R}^{p\times p}$ is a time-varying covariance matrix and $\nu>2$ is the number of degrees of freedom. Note that $p(\mb{y}_t|\mb{V}_t,\nu)$ is a normalized Student $t$ distribution such that $\text{Cov}(\mb{y}_t|\mb{V}_t,\nu)=\mb{V}_t$. Applying the filtering equation (\ref{eq:sdsFgen}) leads to the $t$-GAS model of \cite{GAS2}. Closed form formulas for the score and information matrix are reported in \cite{GAS2}. These authors also proposed two parameterizations of $\mb{V}_t$ leading to positive-definite estimates. The first is similar to the one used in the DCC model of \cite{EngleDCC}, while the second is based on hyperspherical coordinates. In the two parameterizations, the number of time-varying parameters is $k=p+p(p+1)/2$ and $k=p(p+1)/2$, respectively. Unlike the previous examples, the score and the information matrix of this model do not reduce to a compact expression in the observations, since they involve the Jacobian of the chosen parameterization of $\mb{V}_t$ together with duplication, elimination and commutation matrices. To save space, we report the model details and the corresponding recursions in Appendix \ref{app:tgas}.

\subsection{Conditional variances, parameter and filtering uncertainty}
\label{sub:filtIUnc}

Confidence bands can reflect both parameter and filtering uncertainty. Parameter uncertainty arises because the static parameters are replaced by their maximum likelihood estimates, and affects both observation-driven and parameter-driven models; in observation-driven models it can be quantified through the methods developed by \cite{BLASQUES2016875}. Filtering uncertainty arises because the time-varying parameters are not completely revealed by the observations, and is therefore absent from observation-driven models, where they are deterministic functions of past observations. Once such models are regarded as filters, however, one is interested in bands around the filtered and smoothed estimates that reflect the conditional distribution of the underlying state variable.

In linear Gaussian models, filtering uncertainty can be assessed through the variance matrices $\mb{P}_{t+1}$, $\mb{P}_{t|t}$, $\hat{\mb{P}}_t$ introduced in Section \ref{sub:kf}, which provide the conditional variance of the unobserved state variable. It is instead less clear how one can quantify filtering uncertainty in misspecified observation-driven models. \cite{Zamojski} proposed a bootstrap based method for assessing filtering uncertainty in GARCH filters. Confidence bands constructed through this technique tend to underestimate filtering uncertainty, because they are based on bootstraps of the filter rather than the underlying state variable. In addition, the method of \cite{Zamojski} does not allow one to construct out-of-sample confidence bands, which are often needed in practical applications. 

In our framework, in-sample and out-of-sample confidence bands can be constructed by exploiting the relation between Kalman filter recursions and score-driven recursions. In Section \ref{sub:genKf} the steady state variance matrix $\mb{\bar{P}}$ enters the recursions through $\mb{A}=\mb{B}\mb{\bar{P}}$, with the unscaled score, and, $\mb{B}$ being invertible, it can be expressed as $\mb{\bar{P}} = \mb{B}^{-1}\mb{A}$.
In score-driven models featuring a scaling matrix $\mb{S}_t$, the conditional covariance can be defined by matching the two predictive recursions. The driving term is $\mb{B}\mb{P}_t\bs{\nabla}_t$ in Eq. \eqref{eq:kfGeneral3} and $\mb{A}\mb{S}_t\bs{\nabla}_t$ in Eq. \eqref{eq:sdsFgen}, so that the two coincide provided that
\begin{equation}
	\mb{P}_t = \mb{B}^{-1}\mb{A}\mb{S}_t
	\label{eq:PtAlg}
\end{equation}
which we take as the definition of the conditional covariance of a score-driven model.

\begin{remark}
\label{rem:predictive}
The recursions of Section (\ref{sub:genKf}) are written in terms of the predictive density $p(\mb{y}_t|\mb{Y}_{t-1})$, whereas those of Section (\ref{sub:sdsRec}) are run with a conditional density whose parameters are estimated on data. The linear Gaussian case shows that the second is an estimate of the first. Let the data be generated by Eq. \eqref{eq:lgmObs}, \eqref{eq:lgmTrans} with measurement variance $\mb{H}$, and fit to them a score-driven location model with a Gaussian conditional density of unknown variance. Since $y_t$ can be re-written as $y_t=f_t + (\alpha_t-f_t)+\varepsilon_t$ with the latter two terms independent, the maximum likelihood estimate of that variance converges to $\mb{\bar{P}}+\mb{H}=\mb{\bar{F}}$ rather than to $\mb{H}$. What is estimated is therefore the predictive density of the parameter-driven model that generated the data, and not its observation density. Since $\mb{S}_t=\mb{\bar{F}}$, the matrix matching the steady-state Kalman recursion is the gain $\mb{A}=\mb{B}\mb{\bar{P}}\mb{\bar{F}}^{-1}$, and substituting into Eq. \eqref{eq:PtAlg} returns $\mb{P}_t=\mb{\bar{P}}$, as required.
Outside the linear Gaussian case the predictive density is not available in closed form, and the estimated conditional density is to be read as an approximation to it, of the same order as the other approximations involved. 
\end{remark}

Two observations are in order. First, Eq. \eqref{eq:PtAlg} provides an alternative to the recursion in Eq. \eqref{eq:pt1General}. Instead of propagating $\mb{P}_t$ forward through a Riccati-type equation, it recovers the conditional covariance algebraically from the static parameters and the current scaling matrix. In linear Gaussian models the two agree, both returning the steady state matrix $\mb{\bar{P}}$ that solves Eq. \eqref{eq:riccati}. Eq. \eqref{eq:PtAlg} has the practical advantage of not requiring the state noise covariance $\mb{Q}$ to be specified separately, $\mb{Q}$ being in any case recoverable from Eq. \eqref{eq:riccati} whenever it is of interest.
Second, since $\mb{P}_t$ plays the role of a conditional covariance matrix, it must be symmetric and positive semi-definite. This is automatically the case when $k=1$, and more generally whenever $\mb{B}^{-1}\mb{A}$ is symmetric and commutes with $\mb{S}_t$. This holds in particular when $\mb{A}$ and $\mb{B}$ are proportional to the identity matrix, which is the specification most commonly adopted in the score-driven literature. Outside these cases, the right-hand side of Eq. \eqref{eq:PtAlg} should be symmetrized before being used to construct confidence bands.

Substituting Eq. \eqref{eq:PtAlg} into Eq. \eqref{eq:pttGeneral}, the corresponding expression for $\mb{P}_{t|t}$ is:
\begin{equation}
	\mb{P}_{t|t} = \mb{P}_t-\mb{P}_t\bs{\mathcal{I}}_t\mb{P}_t.
	\label{eq:Pt|t}
\end{equation}
Similarly, Eq. \eqref{eq:NtGeneral}, \eqref{eq:hatPtGeneral} specialize to:
\begin{align}
	\mb{N}_{t-1} &= \bs{\mathcal{I}}_t + (\mb{B}-\mb{A}\mb{S}_t\bs{\mathcal{I}}_t)'\mb{N}_t(\mb{B}-\mb{A}\mb{S}_t\bs{\mathcal{I}}_t)\\
	\hat{\mb{P}}_t &= \mb{P}_t-\mb{P}_t\mb{N}_{t-1}\mb{P}_t
	\label{eq:hatPt}
\end{align}
with $\mb{N}_n=\mb{0}$ and $t=n,\dots,1$.

Confidence bands can be computed as quantiles of the conditional distribution of the state variable. For a general state-space model, the latter is non-Gaussian and is not known analytically. 
To construct approximate confidence bands at a given confidence level, one can assume a normal conditional density. The Monte Carlo analysis of Section \ref{sub:mcBands} shows that, for state-space models characterized by a nonlinear non-Gaussian observation density but linear transition equation, this approximation delivers out-of-sample coverage within four percentage points of the nominal level, for all three estimates and at all the confidence levels considered.

As mentioned, when score-driven models are employed as misspecified filters, it is essential to construct confidence bands reflecting both parameter and filtering uncertainty. Following \cite{HAMILTON1986387}, this can be done by adopting the Bayesian perspective that the vector of static parameters, that we denote by $\bs{\theta}$, is a random variable with a certain prior distribution $p(\bs{\theta})$. In practice, $p(\bs{\theta})$ can be set equal to the asymptotic distribution of the maximum likelihood estimate $\tilde{\bs{\theta}}$, as in \cite{BLASQUES2016875}. Let $\mb{f}_t^{\tilde{\bs{\theta}}}$ denote the predictive filter computed from $\tilde{\bs{\theta}}$. It is possible to show (\citealt{HAMILTON1986387}) that the total conditional variance can be decomposed into the sum of two terms:
\begin{equation}
\begin{split}
 \Exp[(\bs{\alpha}_t-\mb{f}_t^{\tilde{\bs{\theta}}})(\bs{\alpha}_t-\mb{f}_t^{\tilde{\bs{\theta}}})'|\mb{Y}_{t-1}] &=\\
 \Exp_{{\bs{\theta}}}[(\bs{\alpha}_t-\mb{f}_t^{\bs{\theta}})(\bs{\alpha}_t- \mb{f}_t^{\bs{\theta}})'|\mb{Y}_{t-1}] &+ \Exp_{{\bs{\theta}}}[(\mb{f}_t^{\bs{\theta}}-\mb{f}_t^{\tilde{\bs{\theta}}})(\mb{f}_t^{\bs{\theta}}-\mb{f}_t^{\tilde{\bs{\theta}}})']=\\
 \Exp_{\bs{\theta}}[\mb{P}_t^{\bs{\theta}}] &+ \Exp_{\bs{\theta}}[(\mb{f}_t^{\bs{\theta}}-\mb{f}_t^{\tilde{\bs{\theta}}})(\mb{f}_t^{\bs{\theta}}-\mb{f}_t^{\tilde{\bs{\theta}}})']
 \end{split}
 \label{eq:filtParUnc}
\end{equation}
where $\Exp_{\bs{\theta}}[\cdot]$ denotes the expectation taken with respect to the prior density $p(\bs{\theta})$. The first term is related to filtering uncertainty. It represents the conditional variance of the state variables averaged over the density $p(\bs{\theta})$. The second term is clearly related to parameter uncertainty, as it represents the variance of $\mb{f}_t^{\bs{\theta}}$ due to the randomness of $\bs{\theta}$ around the maximum likelihood estimate $\tilde{\bs{\theta}}$. Both terms can be computed by simulations, sampling from the prior density $p(\bs{\theta})$ and then taking the sample mean.
The decomposition in Eq. \eqref{eq:filtParUnc} holds exactly provided that the cross term
\begin{equation*}
	\Exp_{\bs{\theta}}\!\left[\Exp[(\bs{\alpha}_t-\mb{f}_t^{\bs{\theta}})|\mb{Y}_{t-1},\bs{\theta}]\,(\mb{f}_t^{\bs{\theta}}-\mb{f}_t^{\tilde{\bs{\theta}}})'\right]
\end{equation*}
vanishes, which is the case when $\mb{f}_t^{\bs{\theta}}$ is the conditional mean of $\bs{\alpha}_t$ given $\mb{Y}_{t-1}$ and $\bs{\theta}$. In our setting $\mb{f}_t^{\bs{\theta}}$ is an approximate filter, so that the cross term does not vanish identically, but is of the same order as the approximation error of the filter itself. The Monte Carlo analysis of Section \ref{sub:mcBands} indicates that neglecting it does not compromise the accuracy of the resulting confidence bands.

We can now make explicit how confidence bands are obtained once an estimate of the conditional variance of the state is available. Let $\bs{\Sigma}_t$ denote the total conditional variance appearing on the left-hand side of Eq. \eqref{eq:filtParUnc}, and let $\alpha$ be the nominal confidence level. Approximating the conditional density of the state by a normal density with mean $\mb{f}_t^{\tilde{\bs{\theta}}}$ and variance $\bs{\Sigma}_t$, the confidence band for the $i$-th component of the state is:
\begin{equation}
	\mb{f}_{t,i}^{\tilde{\bs{\theta}}} \pm z_{(1+\alpha)/2}\sqrt{\bs{\Sigma}_{t,ii}}
	\label{eq:band}
\end{equation}
where $z_{(1+\alpha)/2}$ denotes the corresponding quantile of the standard normal distribution. The same construction applies to the update filter and to the smoother, provided that $(\mb{f}_t,\mb{P}_t,\mb{Y}_{t-1})$ are replaced by $(\mb{f}_{t|t},\mb{P}_{t|t},\mb{Y}_t)$ and by $(\mb{\hat{f}}_t,\hat{\mb{P}}_t,\mb{Y}_n)$, respectively. The two terms entering $\bs{\Sigma}_t$ are computed by simulation, according to the following procedure:
\begin{enumerate}
	\item Estimation of the static parameters, as in Section \ref{sub:sdsRec}, together with the covariance matrix $\bs{\Sigma}_{\bs{\theta}}$ of the estimator, obtained for instance as the inverse of the observed information matrix at the optimum.
	\item Sampling of $J$ values of the static parameters from the prior density:
	\begin{equation*}
		\bs{\theta}^{(j)}\sim N(\tilde{\bs{\theta}},\bs{\Sigma}_{\bs{\theta}}),\qquad j=1,\dots,J
	\end{equation*}
	\item For each draw $\bs{\theta}^{(j)}$, forward and backward recursions of Section \ref{sub:sdsRec}, yielding the estimates $\mb{f}_t^{(j)}$, together with the conditional variances $\mb{P}_t^{(j)}$ obtained from Eq. \eqref{eq:PtAlg}.
	\item Sample counterparts of the two terms of Eq. \eqref{eq:filtParUnc}:
	\begin{align*}
		\bs{\Sigma}_t^{\text{filt}} &= \frac{1}{J}\sum_{j=1}^J\mb{P}_t^{(j)} \\
		\bs{\Sigma}_t^{\text{par}} &= \frac{1}{J}\sum_{j=1}^J(\mb{f}_t^{(j)}-\mb{f}_t^{\tilde{\bs{\theta}}})(\mb{f}_t^{(j)}-\mb{f}_t^{\tilde{\bs{\theta}}})'
	\end{align*}
	\item Confidence band from Eq. \eqref{eq:band}, with $\bs{\Sigma}_t=\bs{\Sigma}_t^{\text{filt}}+\bs{\Sigma}_t^{\text{par}}$.
\end{enumerate}
The decomposition makes it straightforward to isolate the two sources of uncertainty. Setting $\bs{\Sigma}_t=\bs{\Sigma}_t^{\text{filt}}$ delivers bands reflecting filtering uncertainty only, whereas setting $\bs{\Sigma}_t=\bs{\Sigma}_t^{\text{par}}$ delivers bands reflecting parameter uncertainty only. The distinction between in-sample and out-of-sample bands lies entirely in the sample used to obtain $\tilde{\bs{\theta}}$, the recursions being otherwise unchanged. Finally, bands are sometimes needed for a monotone transformation of the state, as when the state is a log-variance and one is interested in the volatility. In such cases it is sufficient to apply the transformation to the endpoints in Eq. \eqref{eq:band}.

The procedure requires $J$ complete forward-backward passes of the filter. Since each pass is linear in the sample size and involves no integration over the latent states, the additional cost remains modest. In simulation-based approaches, by contrast, the same exercise would require nesting the parameter draws within a sequential Monte Carlo scheme.

\section{Monte Carlo analysis}
\label{sec:MonteCarlo}

In this section we examine by Monte Carlo simulations the performance of the recursions of Section \ref{sub:sdsRec}.
As a data generating process, we consider several nonlinear non-Gaussian state-space models for location, scale and duration. To assess the loss of efficiency due to our approximation, we compare the filtered and smoothed estimates recovered by our methodology with those obtained from exact simulation-based methods. To estimate the static parameters of the parameter-driven model we use the ``Numerically Accelerated Importance Sampling'' (NAIS) method of \cite{NAIS}, in a similar fashion to \cite{GAS3}. The smoothed estimate is then computed by importance sampling, as described by \cite{DurbinKoopman}, whereas the predictive and update estimates are obtained from a bootstrap particle filter; the details are discussed in Section \ref{sub:mcExact}.

The DGP considered in this analysis is a general state-space model of the form:
\begin{align}
	{y}_t |{\alpha}_t &\sim p({y}_t|{\alpha}_t;\bs{\theta}) \label{eq:ssm_obs_gen}\\
	\alpha_{t+1} &= {c} + \phi{\alpha}_t + {\eta}_t,\qquad \eta_t\sim \text{N}(0,q) \label{eq:ssm_trans_gen} 
\end{align} 
where ${y}_t$ are the simulated observations and ${\alpha}_t$ the latent state variable that, for simplicity, we set driven by normal innovations $\eta_t$. 

In particular, we concentrate on the following models:
\begin{align*}
\text{Location (Student-\textit{t}):}\quad \quad \quad y_t &= \alpha_t + \eta_t,\quad \eta_t\sim t_{\nu}(0,e^{\lambda})\\
\text{Scale (Gaussian):}\quad \quad\quad y_t &= e^{(\omega+\alpha_t)/2}\epsilon_t,\quad \epsilon_t \sim \text{N}(0,1)\\
\text{Scale (Student-\textit{t}):}\quad\quad \quad y_t &= e^{(\omega+\alpha_t)/2}\epsilon_t,\quad \epsilon_t \sim t_{\nu}(0,1)\\
\text{Duration (Poisson):}\quad \quad\quad y_t &\sim \text{Poiss}(e^{\alpha_t})
\end{align*}
The first model is linear in the location parameter, but has Student-$t$ distributed measurement errors with scale $e^{\lambda}$. The second and third models are both nonlinear stochastic volatility models. They differ because of the observation density, which is Gaussian in the first model and Student-$t$ in the second. The last is a stochastic duration model with a Poisson density. Such models have been applied extensively in the economic and finance literature. Few examples are given by \cite{HarveySV}, \cite{GhyselsHarveyRenault} and \cite{BAUWENS2004381}.

In order to run the filtering algorithm described in Section \ref{sub:sdsRec}, we need to specify the density used to compute the score.
Throughout the analysis we use the density $p(y_t|\alpha_t)|_{f_t}$, with all its parameters estimated jointly with $\bs{\omega}$, $\mb{A}$ and $\mb{B}$; as discussed in Remark \ref{rem:predictive}, estimation is what makes this an approximation to the predictive density of the underlying model. 
\begin{table}[h]
\centering
\setlength{\tabcolsep}{8pt}
\renewcommand{\arraystretch}{1.2}
\begin{tabular}{llcc}
\toprule
 Model  & Distribution & $p(y_t|\alpha_t)|_{f_t}$ & $\bs{\mathcal{I}}_t$\\
\midrule
   Location      & Student-$t$  & $\frac{\Gamma(\frac{\nu+1}{2})}{\Gamma(\frac{\nu}{2})\sqrt{\pi(\nu-2)e^{\lambda}}}\left(1+\frac{(y_t-f_t)^2}{(\nu-2) e^{\lambda}}\right)^{-\frac{\nu+1}{2}}$&  $\frac{\nu(\nu+1)}{(\nu+3)(\nu-2)e^{\lambda}}$\\[6pt]
   Scale & Gaussian & $\frac{1}{\sqrt{2\pi e^{\omega+f_t}}}\exp\left(-\frac{y_t^2}{2e^{\omega+f_t}}\right)$ & $\frac{1}{2}$\\[6pt]
   Scale & Student-$t$ &  $\frac{\Gamma(\frac{\nu+1}{2})}{\Gamma(\frac{\nu}{2})\sqrt{\pi(\nu-2)e^{\omega+f_t}}}\left(1+\frac{y_t^2}{(\nu-2) e^{\omega+f_t}}\right)^{-\frac{\nu+1}{2}}$ & $\frac{\nu}{2(\nu+3)}$\\[6pt]
   Duration  & Poisson & $\frac{e^{y_t f_t} e^{-e^{f_t}}}{y_t!}$ & $e^{f_t}$\\
\bottomrule
  \end{tabular}
\caption{Conditional density and information quantity used in the filtering algorithm for each state-space model. The normalization is $\mb{S}_t=\bs{\mathcal{I}}_t^{-1}$ throughout. }
\label{tab:filterSpec}
\end{table}
We set the scaling matrix equal to the inverse of the information quantity, $\mb{S}_t=\bs{\mathcal{I}}_t^{-1}$, which is the common choice in score-driven models, as in \cite{GAS1} and \cite{Harvey_2013}. The models considered here have a single latent component, so that $k=1$ and the matrix $\mb{P}_t$ defined through Eq. \eqref{eq:PtAlg} is automatically positive, as discussed in Section \ref{sub:filtIUnc}. Table (\ref{tab:filterSpec}) specifies both the densities and the normalization used in the filtering algorithm, whereas Table (\ref{tab:statPar}) reports the values of the static parameters of the four state-space models. 
\begin{table}[h]
\centering
\setlength{\tabcolsep}{8pt}
\renewcommand{\arraystretch}{1.1}
\begin{tabular}{lcccccc}
\toprule
 Model  & $c$ & $\phi$ & $q$ & $\lambda$ & $\omega$ & $\nu$\\
\midrule
 Location (Student-$t$) & 0.01 & 0.98  & 0.01 & 0.01 & --- & 5\\
 Scale (Gaussian) & 0 & 0.98 & 0.01 & --- & 0.1 & ---\\
 Scale (Student-$t$) & 0 & 0.98 & 0.01 & --- & 0.1 & 5\\
 Duration (Poisson) & 0.001 & 0.98 & 0.01 & --- & --- & ---\\
\bottomrule
  \end{tabular}
 \caption{Values of the static parameters used to simulate the four state-space models used in the analysis.}
\label{tab:statPar}
\end{table}

The Monte Carlo study is based on 1000 replications of $n=4000$ observations of the state-space models described above. Each sample is divided into two sub-samples of equal size. The first sub-sample is used to estimate the models, whereas the second is used to compute the mean square error (MSE) of the three estimates, so that all comparisons reported below are out of sample. The static parameters are estimated by maximizing numerically the approximate log-likelihood $\hat{L}=\sum_{t=1}^n \log p(y_t|\alpha_t)|_{f_t}$.

Since the predictive filter, the update filter and the smoother are computed on the same data, their MSE are strongly dependent across replications, and the marginal averages alone give a misleading impression of the precision of the comparison. We therefore compute the relative gains \emph{replication by replication} and average them afterwards. This paired form removes the variability of the level of the MSE across replications, which is an order of magnitude larger than the differences of interest, and delivers Monte Carlo standard errors that are two orders of magnitude smaller than those of the levels.\medskip

\begin{table}[htbp]
\centering
\caption{Out-of-sample mean square error of the score-driven predictive filter (SDF), update filter (SDU) and smoother (SDS).}
\label{tab:MCsd}
\begin{tabular}{lccc c ccc}
\toprule
 & \multicolumn{3}{c}{Mean square error} & & \multicolumn{3}{c}{Relative gain (\%)} \\
\cmidrule(lr){2-4} \cmidrule(lr){6-8}
Model & SDF & SDU & SDS & & SDU/SDF & SDS/SDF & SDS/SDU \\
\midrule
Location ($t$) & 0.0824 & 0.0751 & 0.0459 & & $-8.81$ & $-44.05$ & $-38.67$ \\
 & & & & & \scriptsize{(0.03)} & \scriptsize{(0.15)} & \scriptsize{(0.16)} \\
Scale (Gaussian) & 0.1231 & 0.1173 & 0.0778 & & $-4.74$ & $-36.34$ & $-33.20$ \\
 & & & & & \scriptsize{(0.02)} & \scriptsize{(0.21)} & \scriptsize{(0.21)} \\
Scale ($t$) & 0.1448 & 0.1397 & 0.0982 & & $-3.52$ & $-31.82$ & $-29.37$ \\
 & & & & & \scriptsize{(0.02)} & \scriptsize{(0.24)} & \scriptsize{(0.24)} \\
Duration (Poisson) & 0.0906 & 0.0836 & 0.0529 & & $-7.72$ & $-41.45$ & $-36.57$ \\
 & & & & & \scriptsize{(0.03)} & \scriptsize{(0.17)} & \scriptsize{(0.17)} \\
\bottomrule
\end{tabular}
\begin{minipage}{0.92\textwidth}\vspace{2mm}\footnotesize
Notes: 1000 replications of $n=4000$ observations. The static parameters are estimated on the first half of each sample and the three estimates are compared with the true signal on the second half, so that all comparisons are out of sample. Relative gains are computed replication by replication and then averaged; since the three estimates are obtained from the same data, this paired form removes the across-replication variability of the level of the mean square error. Monte Carlo standard errors of the paired gains are reported in parentheses.
\end{minipage}
\end{table}

\begin{figure}[htbp]
\centering
  \includegraphics[width=\textwidth]{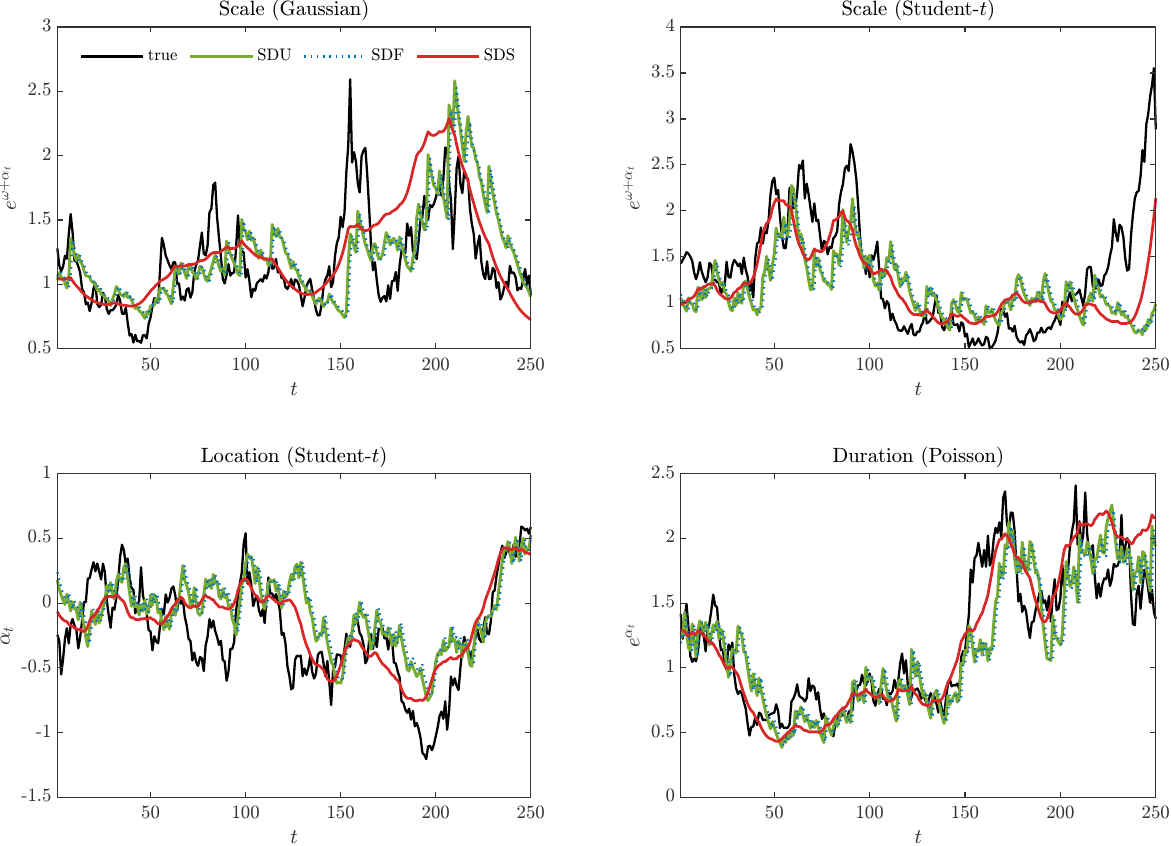}
  \caption{For one particular simulation, out-of-sample estimates of the signal obtained from the score-driven predictive filter (SDF), update filter (SDU) and smoother (SDS), together with the true signal. The two filters are nearly indistinguishable from one another, whereas the smoother tracks the latent path visibly more closely. For the two stochastic volatility models the conditional variance $e^{\omega+\alpha_t}$ is shown, and for the duration model the intensity $e^{\alpha_t}$.}
  \label{fig:sdCompare}
\end{figure}

\begin{figure}[htbp]
\centering
  \includegraphics[width=\textwidth]{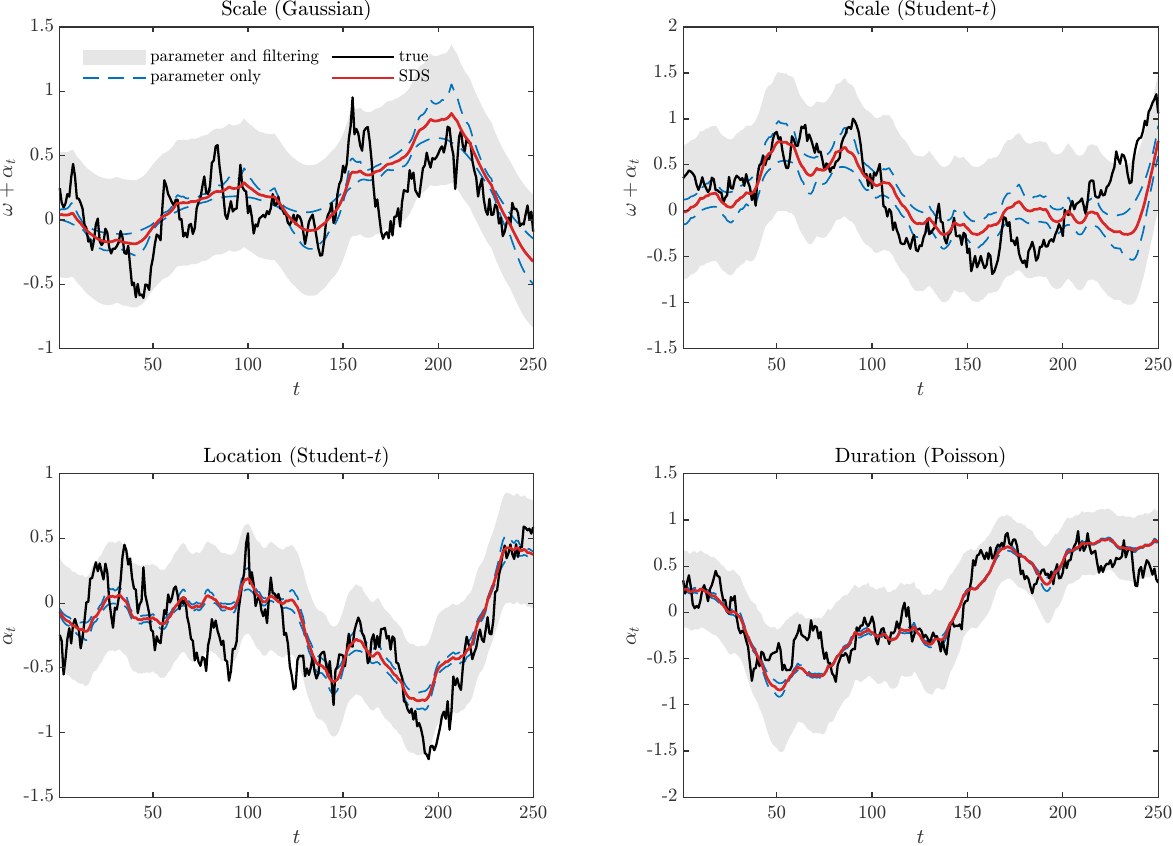}
  \caption{Out-of-sample confidence bands at the $95\%$ nominal level around the smoothed estimates, for one particular simulation. The shaded area is obtained from the sum of the filtering and the parameter components of the conditional variance, as in Eq. \eqref{eq:band}; the dashed lines delimit the band that would be obtained from the parameter component alone. Neglecting filtering uncertainty produces bands that are far too narrow.}
  \label{fig:sdBands}
\end{figure}

Table (\ref{tab:MCsd}) reports the results. Three regularities emerge, and they are the same across the four models.
First, exploiting the contemporaneous observation is worth having but is not where the bulk of the gain lies. Moving from the predictive to the update filter lowers the MSE by between $3.5\%$ and $8.8\%$, depending on the model. Second, exploiting the whole sample changes the picture entirely. The smoother lowers the MSE by between $32\%$ and $44\%$ relative to the predictive filter, and by between $29\%$ and $39\%$ relative to the update filter. The gain from smoothing is therefore roughly six times the gain from updating. Third, the ordering is not merely an average property. The inequality $\text{MSE}(\text{SDS})\le\text{MSE}(\text{SDU})\le\text{MSE}(\text{SDF})$ holds in \emph{every one} of the 1000 replications, for all four models.

The magnitude of the gain varies systematically with how informative a single observation is about the signal, and does so monotonically across the four specifications. It is largest for the location model with Student-$t$ errors and for the Poisson duration model, and smallest for the scale model with Student-$t$ errors, where the conditional information $\bs{\mathcal{I}}_t=\nu/(2(\nu+3))$ is smallest among the four. The predictive filter is the only one of the three that cannot use $y_t$, and is therefore left with the state innovation however precise the observations are, whereas the update filter and the smoother do exploit it. The more each observation reveals, the wider the resulting gap.

Figure (\ref{fig:sdCompare}) shows the same result on a single replication. The two filters are visually almost indistinguishable, which is the graphical counterpart of the modest gain from updating, whereas the smoother is markedly less erratic and tracks the latent path more closely.

\subsection{Confidence bands}
\label{sub:mcBands}

We next assess the confidence bands of Section \ref{sub:filtIUnc}. Bands are constructed out of sample following the procedure described there, with $J=200$ parameter draws per replication and $\bs{\Sigma}_{\bs{\theta}}$ set equal to the inverse of the observed information at the optimum, on 250 replications. Three nominal levels are considered, $\alpha=0.90,0.95,0.99$. Table (\ref{tab:MCbands}) reports the average fraction of out-of-sample periods in which the true signal falls inside the band.

\begin{table}[htbp]
\centering
\caption{Average out-of-sample coverage of the confidence bands.}
\label{tab:MCbands}
\begin{tabular}{llcccc}
\toprule
& & Location ($t$) & Scale (Gaussian) & Scale ($t$) & Duration (Poisson) \\
\midrule
\multicolumn{6}{l}{\textit{Nominal level $\alpha=0.90$}} \\[2pt]
\multirow{3}{*}{Par.\ \& filt.} & SDF & 0.8979 & 0.8638 & 0.8955 & 0.8670 \\
 & SDU & 0.8981 & 0.8635 & 0.8946 & 0.8672 \\
 & SDS & 0.9002 & 0.8662 & 0.8939 & 0.8649 \\
\addlinespace[2pt]
 Par.\ only & SDF & 0.1690 & 0.2000 & 0.2990 & 0.1574 \\
\midrule
\multicolumn{6}{l}{\textit{Nominal level $\alpha=0.95$}} \\[2pt]
\multirow{3}{*}{Par.\ \& filt.} & SDF & 0.9476 & 0.9231 & 0.9454 & 0.9230 \\
 & SDU & 0.9477 & 0.9228 & 0.9453 & 0.9236 \\
 & SDS & 0.9491 & 0.9252 & 0.9450 & 0.9227 \\
\addlinespace[2pt]
 Par.\ only & SDF & 0.2007 & 0.2370 & 0.3512 & 0.1868 \\
\midrule
\multicolumn{6}{l}{\textit{Nominal level $\alpha=0.99$}} \\[2pt]
\multirow{3}{*}{Par.\ \& filt.} & SDF & 0.9884 & 0.9785 & 0.9868 & 0.9773 \\
 & SDU & 0.9884 & 0.9786 & 0.9867 & 0.9775 \\
 & SDS & 0.9888 & 0.9797 & 0.9873 & 0.9773 \\
\addlinespace[2pt]
 Par.\ only & SDF & 0.2612 & 0.3074 & 0.4481 & 0.2433 \\
\bottomrule
\end{tabular}
\begin{minipage}{0.92\textwidth}\vspace{2mm}\footnotesize
Notes: 250 replications of $n=4000$ observations, bands constructed out of sample with $J=200$ parameter draws per replication following the procedure of Section \ref{sub:filtIUnc}. The upper block reports the coverage of bands built on the sum of the filtering and the parameter components of the conditional variance; the last row of each block reports the coverage obtained from the parameter component alone. A well calibrated band has average coverage equal to the nominal level.
\end{minipage}
\end{table}

The bands are reasonably well calibrated, and the quality of the calibration varies across models in an interpretable way. For the location model with Student-$t$ errors the match is essentially exact, the coverage departing from the nominal level by at most a quarter of a percentage point at all three levels; the scale model with Student-$t$ errors is almost as accurate, within six tenths of a point. For the Gaussian scale model and for the Poisson duration model the bands are somewhat too narrow, undercovering by about three and a half percentage points at the $90\%$ level, by less than three at the $95\%$ level and by little more than one at the $99\%$ level. The residual undercoverage is therefore concentrated in the centre of the distribution rather than in its tails, which is the pattern produced by a conditional variance that is slightly too small rather than by a misspecified shape. Two further features are worth noting. The first is that the calibration is essentially identical for the predictive filter, the update filter and the smoother, the three coverage rates never differing by more than a few tenths of a percentage point. Since the three variances are delivered by three different recursions, their agreement is a non-trivial check on the identification in Eq. \eqref{eq:PtAlg}--\eqref{eq:hatPt}. The second is that the positivity requirement discussed after Eq. \eqref{eq:PtAlg} is never binding. Not a single parameter draw had to be discarded in any replication.

The last row of each panel shows what happens when filtering uncertainty is ignored and the band is built on the parameter component alone. The coverage collapses to between $0.16$ and $0.30$ against a nominal level of $0.90$, so that such bands capture less than a third of the uncertainty they are meant to represent. The reason is a difference in the order of magnitude: parameter uncertainty is $O(n^{-1})$ and vanishes as the estimation sample grows, whereas filtering uncertainty is $O(1)$ and does not vanish at all, because the state is never revealed by the observations. Figure (\ref{fig:sdBands}) makes the point graphically on a single replication. Accounting for filtering uncertainty is therefore not a refinement but a precondition for computing meaningful confidence bands around score-driven estimates.

\subsection{Comparison with the exact method}
\label{sub:mcExact}

Although characterising the approximation error is not the purpose of this analysis, it is useful to record its order of magnitude. We therefore compare the score-driven estimates with those of an exact simulation-based method on 50 of the replications above. The exact method combines a bootstrap particle filter for the predictive and update estimates with importance sampling for the smoothed estimate. The particle filter is used for the two filtered quantities because they condition on an expanding information set, which a single full-sample importance density cannot deliver without being rebuilt at every $t$. In its bootstrap form the particles are propagated through the state transition and weighted by the observation density. With a single latent state, ten thousand particles and systematic resampling at every step, the weights do not degenerate, so that an adapted proposal would improve efficiency without changing the benchmark. For the three models in which the state enters the scale or the intensity we use the Gaussian importance density of \cite{NAIS}; for the Student-$t$ location model that density degenerates, and we rely instead on the scale-mixture representation of the $t$ distribution, which handles the tails exactly.\footnote{In the location model the observation log-density is a Student-$t$ log-density in the state, so its curvature is negative at every outlier and a Gaussian importance density cannot match its tails. The resulting effective sample size is of order one out of four hundred.} Since the two blocks are run on the same samples, the reported losses are mean paired differences.

\begin{table}[htbp]
\centering
\caption{Out-of-sample mean square error: comparison with the exact simulation-based method.}
\label{tab:MCref}
\begin{tabular}{lccc c ccc}
\toprule
 & \multicolumn{3}{c}{Exact method (MSE)} & & \multicolumn{3}{c}{Score-driven MSE loss (\%)} \\
\cmidrule(lr){2-4} \cmidrule(lr){6-8}
Model & Pred. & Upd. & Smooth & & SDF & SDU & SDS \\
\midrule
Location ($t$) & 0.0809 & 0.0737 & 0.0459 & & $-0.25$ & $-0.13$ & $-1.41$ \\
Scale (Gaussian) & 0.1182 & 0.1121 & 0.0718 & & $+5.13$ & $+5.46$ & $+7.04$ \\
Scale ($t$) & 0.1364 & 0.1316 & 0.0910 & & $+5.38$ & $+5.57$ & $+7.79$ \\
Duration (Poisson) & 0.0853 & 0.0784 & 0.0499 & & $+5.64$ & $+6.23$ & $+7.09$ \\
\bottomrule
\end{tabular}
\begin{minipage}{0.92\textwidth}\vspace{2mm}\footnotesize
Notes: 50 replications of $n=4000$ observations. The first three columns report the out-of-sample mean square error of the exact method; the last three report the percentage increase in mean square error of the corresponding score-driven estimate, so that a positive entry means that the score-driven estimate is the less accurate of the two. The exact method combines a bootstrap particle filter for the two filtered quantities with importance sampling for the smoothed one. They are recomputed on the same 50 samples, so that the reported losses are mean paired differences and are not contaminated by the variability of the level of the mean square error across replications. For the location model the importance sampler has an effective sample size of the order of ten out of two hundred, and its own Monte Carlo error is of the same order as the differences reported here; the corresponding entries should be read as indicating that the two methods cannot be told apart on that model, rather than as evidence that one dominates.
\end{minipage}
\end{table}

For the two scale models and the duration model the score-driven estimates lose between $5\%$ and $8\%$ in mean square error, with the loss slightly larger for the smoother than for the two filters. For the location model the entries are instead slightly negative, which taken literally would mean that the approximation outperforms the exact method. The benchmark is itself imprecise on this model, since the scale-mixture sampler retains an effective sample size of about ten draws out of two hundred, so that its own Monte Carlo error is as large as the differences being measured. Therefore, the two methods cannot be told apart here.

The exact method is also considerably more expensive, but the gap lies entirely in the estimation of the static parameters, the score-driven likelihood being available in closed form while each evaluation of the simulated likelihood requires several hundred importance paths. The filtering and smoothing recursions themselves are negligible in both approaches. Appendix \ref{app:cost} documents this.

\section{Empirical illustration}
\label{sec:empirics}  

It is interesting to investigate whether the results found in the simulation study of Section \ref{sec:MonteCarlo} also hold on empirical data. In particular, we aim to provide a quantitative assessment of the improvement of the update filter and of the smoother over the predictive filter in a problem of empirical relevance. Unlike the simulation study, it is generally difficult to perform such analysis empirically, given that time-varying parameters do not belong to the econometrician’s information set and cannot be employed as a benchmark in the loss function. However, when dealing with conditional covariance estimates computed from \textit{daily} log-returns, one can use realized covariance computed from \textit{intraday} log-returns as an accurate proxy of the true latent covariance (\citealt{AndersenBollerslev}). Loss functions can therefore be built as if covariances were observed. This empirical analysis will also show the advantages of the SDS in a highly multivariate framework, where the use of simulation-based methods is computationally problematic or even unfeasible; we quantify the cost of the two approaches in Appendix \ref{app:cost}.

Our dataset consists of unbalanced 1-minute transaction data of Standard \& Poor's 500 constituents over the period from 03-01-2005 to 31-08-2025. The analysis is performed on the subsample comprising the last $T=2000$ days, in order to avoid discontinuities due to changes on index composition. We consider trades from 9:30 to 16:00, leading to 390 timestamps per day. We define the liquidity of an asset as one minus the fraction of missing 1-minute prices, and exclude from the sample the assets whose liquidity is smaller than 98\%. As a final outcome of our filtering procedure, we obtain $N=191$ assets.

In order to estimate conditional covariances from daily log-returns, we use the $t$-GAS model of \cite{GAS2} described in section \ref{sec:Examples}. Compared to standard conditional covariance models, the main advantage of the $t$-GAS is that it updates covariances by taking into account the full shape of the Student $t$ observation density and thus providing robustness against outliers (see discussions on \citealt{GAS2}). We implement the parameterization based on hyperspherical coordinates, as it generally leads to better estimates. The number of time-varying parameters grows as $p^2$, where $p$ is the number of assets. 

Among the universe of $N=191$ assets, we select random groups of $p=5,10,20$ assets. In particular, for each cross-section dimension $p$, we randomly choose four groups. The analysis is thus performed on 12 different groups of assets. As done in the simulation study, we use the RMSE and Qlike as loss functions. The benchmark used in the loss function is the realized covariance estimator of \cite{BNSmulti} computed at the 5-minutes sampling frequency. The use of other estimators does not alter the outcome of the experiment. For each group of assets, the $t$-GAS is estimated on the time-series of $T=2000$ open-to-close log-returns. We thus compute (i) the predictive filter $\mb{f}_t$ (SDF), (ii) the update filter $\mb{f}_{t|t}$ (SDU) and (iii) the smoother $\hat{\mb{f}}_t$ (SDS). The statistical significance of loss differences is tested through the model confidence set of \cite{MCS} at the 90\% confidence level.

Tables \ref{tab:emp_p=5}, \ref{tab:emp_p=10}, \ref{tab:emp_p=20} show the results of the analysis, for $p=5$, $10$, $20$, respectively. We first note that covariance estimates constructed through the predictive and update filters feature worse RMSE and Qlike and are always excluded from the model confidence set. The SDF is the least informative of the three, since only past log-returns are used when reconstructing time-varying parameters. Leveraging the contemporaneous observation is not sufficient to include the SDU in the model confidence set. The smoothed estimates provide a better reconstruction of realized covariance, suggesting that future observations contain relevant information on today's covariance. Note also that relative gains are similar across different dimensions, meaning that the SDS is not affected by the proliferation of time-varying parameters when the number of assets increases.

The above results suggest that, when extracting latent covariance, the smoothing provided by the $t$-GAS model is effective in aggregating all available information. Compared to standard score-driven filtered estimates, the update filter $\mb{f}_{t|t}$ and the smoother $\hat{\mb{f}}_t$ can thus be regarded as providing a more accurate estimate of latent covariance. As seen in the simulation study, this is true for a large class of dynamic models. We therefore advocate the use of the SDS in place of standard filtered estimates in signal reconstruction analysis.


\begin{table}[htbp]
\centering
\setlength{\tabcolsep}{14pt}
\begin{tabular}{lcccc}
\toprule
  & Portfolio 1 & Portfolio 2 & Portfolio 3 & Portfolio 4 \\
\midrule
\multicolumn{5}{l}{\textit{RMSE} $\times 10^{6}$} \\[2pt]
\multirow{2}{*}{SDF} & 0.1322      & 0.1134      & 0.2614      & 0.1946 \\
                     & 1.0000      & 1.0000      & 1.0000      & 1.0000 \\
\addlinespace[3pt]
\multirow{2}{*}{SDU} & 0.1311      & 0.1124      & 0.2572      & 0.1924 \\
                     & 0.9920      & 0.9911      & 0.9838      & 0.9890 \\
\addlinespace[3pt]
\multirow{2}{*}{SDS} & 0.1225$^*$  & 0.1072$^*$  & 0.2314$^*$  & 0.1817$^*$ \\
                     & 0.9270      & 0.9453      & 0.8851      & 0.9340 \\
\midrule
\multicolumn{5}{l}{\textit{Qlike}} \\[2pt]
\multirow{2}{*}{SDF} & $-36.9916$     & $-36.9871$     & $-35.6391$     & $-36.2615$ \\
                     & 1.0000         & 1.0000         & 1.0000         & 1.0000 \\
\addlinespace[3pt]
\multirow{2}{*}{SDU} & $-37.0797$     & $-37.0597$     & $-35.7117$     & $-36.3487$ \\
                     & 0.9976         & 0.9980         & 0.9980         & 0.9976 \\
\addlinespace[3pt]
\multirow{2}{*}{SDS} & $-37.2402^*$   & $-37.2133^*$   & $-35.8402^*$   & $-36.4976^*$ \\
                     & 0.9933         & 0.9939         & 0.9944         & 0.9935 \\
\bottomrule
   \end{tabular}
\caption{Absolute and relative RMSE and Qlike of the SDF, SDU and SDS estimates of the $t$-GAS model, for the four randomly selected portfolios with $p=5$ assets. For each estimator the first row reports the loss and the second its ratio to the loss of the SDF. An asterisk indicates that the estimator belongs to the model confidence set at the $90\%$ confidence level.}
\label{tab:emp_p=5}
\end{table}


\begin{table}[h!]
\centering
\normalsize
\setlength{\tabcolsep}{14pt}
\begin{tabular}{lcccc}
\toprule
  & Portfolio 1 & Portfolio 2 & Portfolio 3 & Portfolio 4 \\
\midrule
\multicolumn{5}{l}{\textit{RMSE} $\times 10^{6}$} \\[2pt]
\multirow{2}{*}{SDF} & 0.1320      & 0.1461      & 0.1663      & 0.0922 \\
                     & 1.0000      & 1.0000      & 1.0000      & 1.0000 \\
\addlinespace[3pt]
\multirow{2}{*}{SDU} & 0.1316      & 0.1459      & 0.1656      & 0.0920 \\
                     & 0.9966      & 0.9988      & 0.9960      & 0.9964 \\
\addlinespace[3pt]
\multirow{2}{*}{SDS} & 0.1264$^*$  & 0.1445$^*$  & 0.1624$^*$  & 0.0891$^*$ \\
                     & 0.9576      & 0.9891      & 0.9765      & 0.9656 \\
\midrule
\multicolumn{5}{l}{\textit{Qlike}} \\[2pt]
\multirow{2}{*}{SDF} & $-75.2130$   & $-76.0316$   & $-74.7989$   & $-74.0731$ \\
                     & 1.0000       & 1.0000       & 1.0000       & 1.0000 \\
\addlinespace[3pt]
\multirow{2}{*}{SDU} & $-75.3272$   & $-76.1567$   & $-74.9519$   & $-74.1858$ \\
                     & 0.9985       & 0.9983       & 0.9980       & 0.9985 \\
\addlinespace[3pt]
\multirow{2}{*}{SDS} & $-75.5944^*$ & $-76.3513^*$ & $-75.1788^*$ & $-74.3785^*$ \\
                     & 0.9949       & 0.9958       & 0.9949       & 0.9959 \\
\bottomrule
   \end{tabular}
\caption{Absolute and relative RMSE and Qlike of the SDF, SDU and SDS estimates of the $t$-GAS model, for the four randomly selected portfolios with $p=10$ assets. For each estimator the first row reports the loss and the second its ratio to the loss of the SDF. An asterisk indicates that the estimator belongs to the model confidence set at the $90\%$ confidence level.}
\label{tab:emp_p=10}
\end{table}


\begin{table}[h!]
\centering
\normalsize
\setlength{\tabcolsep}{14pt}
\begin{tabular}{lcccc}
\toprule
  & Portfolio 1 & Portfolio 2 & Portfolio 3 & Portfolio 4 \\
\midrule
\multicolumn{5}{l}{\textit{RMSE} $\times 10^{6}$} \\[2pt]
\multirow{2}{*}{SDF} & 0.1224      & 0.1302      & 0.0736      & 0.0777 \\
                     & 1.0000      & 1.0000      & 1.0000      & 1.0000 \\
\addlinespace[3pt]
\multirow{2}{*}{SDU} & 0.1222      & 0.1283      & 0.0734      & 0.0776 \\
                     & 0.9982      & 0.9851      & 0.9970      & 0.9986 \\
\addlinespace[3pt]
\multirow{2}{*}{SDS} & 0.1196$^*$  & 0.1273$^*$  & 0.0728$^*$  & 0.0771$^*$ \\
                     & 0.9771      & 0.9781      & 0.9899      & 0.9927 \\
\midrule
\multicolumn{5}{l}{\textit{Qlike}} \\[2pt]
\multirow{2}{*}{SDF} & $-149.1254$   & $-146.6747$   & $-147.1636$   & $-151.8513$ \\
                     & 1.0000        & 1.0000        & 1.0000        & 1.0000 \\
\addlinespace[3pt]
\multirow{2}{*}{SDU} & $-149.3413$   & $-146.9650$   & $-147.4905$   & $-152.0304$ \\
                     & 0.9985        & 0.9980        & 0.9978        & 0.9988 \\
\addlinespace[3pt]
\multirow{2}{*}{SDS} & $-149.7258^*$ & $-147.4616^*$ & $-147.6880^*$ & $-152.3787^*$ \\
                     & 0.9960        & 0.9947        & 0.9964        & 0.9965 \\
\bottomrule
   \end{tabular}
\caption{Absolute and relative RMSE and Qlike of the SDF, SDU and SDS estimates of the $t$-GAS model, for the four randomly selected portfolios with $p=20$ assets. For each estimator the first row reports the loss and the second its ratio to the loss of the SDF. An asterisk indicates that the estimator belongs to the model confidence set at the $90\%$ confidence level.}
\label{tab:emp_p=20}
\end{table}

\section{Conclusions}
\label{sec:Conclusions}

Correctly specified observation-driven models are purely predictive, meaning that past observations include all the relevant information related to the dynamics of the time-varying parameters. As such, there is no room for smoothing and the only form of uncertainty is that coming from replacing the true static parameters with their maximum-likelihood estimates.

In this paper we adopt a different view and assume that observation-driven models are misspecified filters, meaning that data are generated by a different dynamic specification, typically a nonlinear non-Gaussian state-space model. In this framework, the time-varying parameters are not completely revealed by past observations, and thus one needs a methodology for smoothing and for assessing filtering uncertainty.   

Starting from a general representation of the Kalman filter and smoothing recursions in terms of the score of the conditional density, we propose new recursions providing
the update filter and the smoother for score-driven models, together with the conditional variances attached to the three estimates. What makes this possible is that the predictive step of the Kalman filter, once written in terms of the score, is itself a score-driven recursion. The remaining Kalman recursions can therefore be carried over by the same route, replacing the Gaussian score with the score of the observation density at hand. From the conditional variances we further obtain in-sample and out-of-sample confidence bands accounting for both parameter and filtering uncertainty.

The simulation study quantifies what is gained by moving beyond the predictive filter. Exploiting the contemporaneous observation lowers the mean square error by $3.5\%$ to $8.8\%$; exploiting the whole sample lowers it by $32\%$ to $44\%$, so that the information contained in the future of the sample is worth roughly six times the information contained in the present. The ordering of the three estimates holds in every one of the thousand replications and for every model considered, and the size of the gain varies systematically with how informative a single observation is about the state, growing with it, since the predictive filter is the only estimate that cannot exploit the contemporaneous observation.

Filtering uncertainty plays a key role when computing confidence bands in misspecified score-driven models. Bands built on the conditional variances delivered by our recursions attain their nominal coverage, the discrepancy never exceeding four percentage points across models, estimators and confidence levels, and staying below one point for two of the four models. Bands that account for parameter uncertainty alone, by contrast, capture less than a third of the uncertainty they are meant to represent. Parameter uncertainty is of order $n^{-1}$ and vanishes as the estimation sample grows, whereas filtering uncertainty does not vanish at all, because the state is never revealed by the observations.

These results have relevant implications when applying observation-driven models to real data. If the latter are better described by a nonlinear non-Gaussian model, the use of correctly specified observation-driven models may be too restrictive, as one cannot exploit the information of contemporaneous and future observations to estimate the time-varying parameters. Furthermore, it leads to narrow confidence bands, which may have deep policy and institutional implications.
Our empirical illustration confirms the simulation evidence on a large cross-section of US equities: when the target is the latent covariance of daily returns, the smoothed estimates are the only ones to survive the model confidence set, and the gain is stable as the number of assets grows. This last point matters in practice, since it is precisely in highly multivariate settings that simulation-based alternatives become problematic.

Two directions seem worth pursuing. The first concerns models with several latent components, to which the recursions carry over unchanged, although the identification of the conditional covariance then requires more care, as discussed in Appendix \ref{app:ident}. The second concerns the static parameters, which we have treated as estimated once and for all; nothing prevents the same construction from being applied recursively, which would be the natural route to a real-time implementation.

\clearpage

\appendix
\appendixpageoff
\numberwithin{equation}{section}

\huge{\textbf{Appendix}}
\normalsize
\vspace{0.5cm}

\section{Computational cost}
\label{app:cost}

The score-driven and the simulation-based approaches differ in computational cost by two orders of magnitude. It is worth being precise about where the difference comes from. It lies almost entirely in the estimation of the static parameters, and hardly at all in filtering and smoothing.

Table \ref{tab:cost} reports the cost of one replication of the Monte Carlo experiment, separately for the three stages.

\begin{table}[htbp]
\centering
\caption{Computational cost of one replication, in seconds, for $n=4000$ observations.}
\label{tab:cost}
\begin{tabular}{lccc c ccc c c}
\toprule
 & \multicolumn{3}{c}{Score-driven} & & \multicolumn{3}{c}{Simulation-based} & & \\
\cmidrule(lr){2-4} \cmidrule(lr){6-8}
Model & Est. & Filt. & Smooth & & Est. & Filt. & Smooth & & Ratio \\
\midrule
Location ($t$)     & 0.879 & 0.0020 & 0.00007 & & 115.4 & 1.46 & 0.29 & & 133 \\
Scale (Gaussian)   & 0.149 & 0.0008 & 0.00000 & &  16.1 & 1.43 & 0.23 & & 118 \\
Scale ($t$)        & 0.731 & 0.0025 & 0.00002 & &  23.6 & 1.71 & 0.25 & &  35 \\
Duration (Poisson) & 0.203 & 0.0010 & 0.00000 & &  16.4 & 1.35 & 0.24 & &  88 \\
\bottomrule
\end{tabular}
\begin{minipage}{0.92\textwidth}\vspace{2mm}\footnotesize
Notes: median over five replications. ``Est.'' is the numerical maximization of
the log-likelihood on the first half of the sample, ``Filt.'' one pass of the
predictive and update filters over the whole sample, and ``Smooth'' the
additional backward pass. For the simulation-based method the filtered
quantities are obtained from a bootstrap particle filter with $10^{4}$ particles
and the smoothed one from importance sampling with $400$ paths. The last column
is the ratio between the two totals.
\end{minipage}
\end{table}

Two things stand out. First, within each method the recursions themselves are negligible. The score-driven forward pass takes between one and two milliseconds on a sample of four thousand observations, and the backward pass adds between $0.3\%$ and $1.8\%$ of that, because it evaluates no density at all. It recycles the scores and the information quantities already computed and stored going forward. Adding the conditional variances of Eq. \eqref{eq:Pt|t} and \eqref{eq:hatPt} leaves the order of magnitude unchanged. The claim that smoothing costs one additional backward pass is therefore not merely qualitative, since that additional pass is in practice free.

Second, the entire gap between the two approaches is an estimation gap. The reason is structural rather than implementational. In score-driven models the likelihood is available in closed form, so one evaluation of the objective costs a single $O(n)$ recursion. In the parameter-driven formulation the likelihood is an integral over the whole path of the state, and each evaluation requires $N$ importance paths, every one of which involves a Kalman filter and smoother pass over the sample; with $N=400$ this is some two to three orders of magnitude more expensive per evaluation, and the optimizer requires a comparable number of evaluations in the two cases. The ratio of total times, between $35$ and $133$ in our experiment, is essentially the ratio of the two estimation costs.

This also explains why the ratio varies so much across models. It is largest for the Student-$t$ location model, where five static parameters have to be estimated and each simulated likelihood evaluation is correspondingly more expensive, and smallest for the scale model with Student-$t$ errors, where the score-driven objective is itself comparatively slow to optimize. The variation reflects the difficulty of the optimization problem, not a property of the filtering recursions.

A caveat on how these numbers should be read. They compare two particular implementations, and the absolute level of either column could be moved appreciably by tuning, for instance through a smaller number of importance paths, analytical derivatives or a different optimizer. What is robust is the decomposition. In both methods the cost is concentrated in estimation, and in both methods smoothing is essentially free once filtering has been carried out.

\section{Proof of Proposition \ref{prop:steadyState}}
\label{app:prop}

We first compute the score of the conditional log-likelihood (\ref{eq:ll}) with respect to the predictive filter $\mb{a}_t$. Since $\mb{v}_t=\mb{y}_t-\mb{Z}\mb{a}_t$ and $\mb{F}_t$ does not depend on $\mb{a}_t$, the chain rule gives:
\begin{equation}
	\bs{\nabla}_t=\left[\frac{\partial\log p(\mb{y}_t|\mb{Y}_{t-1})}{\partial \mb{a}_t'}\right]'=\left[\frac{\partial\log p(\mb{y}_t|\mb{Y}_{t-1})}{\partial \mb{v}_t'}\frac{\partial \mb{v}_t}{\partial \mb{a}_t'}\right]'=\left[\mb{v}_t'\mb{F}_t^{-1}\mb{Z}\right]'=\mb{Z}'\mb{F}_t^{-1}\mb{v}_t
	\label{eq:app:score}
\end{equation}
where the two minus signs arising from $\partial\log p(\mb{y}_t|\mb{Y}_{t-1})/\partial \mb{v}_t'=-\mb{v}_t'\mb{F}_t^{-1}$ and $\partial \mb{v}_t/\partial \mb{a}_t'=-\mb{Z}$ cancel out. The conditional Fisher information matrix follows as:
\begin{equation}
	\bs{\mathcal{I}}_t=\Exp[\bs{\nabla}_t\bs{\nabla}_t'|\mb{Y}_{t-1}]=\mb{Z}'\mb{F}_t^{-1}\Exp[\mb{v}_t\mb{v}_t'|\mb{Y}_{t-1}]\mb{F}_t^{-1}\mb{Z}=\mb{Z}'\mb{F}_t^{-1}\mb{Z}
	\label{eq:app:info}
\end{equation}
where we used the fact that $\Exp[\mb{v}_t\mb{v}_t'|\mb{Y}_{t-1}]=\mb{F}_t$. Substituting (\ref{eq:app:score}) into the update and prediction steps (\ref{eq:kfStandard2}), (\ref{eq:kfStandard3}) of the Kalman filter, we obtain:
\begin{align}
	\mb{a}_{t|t} &= \mb{a}_t+\mb{P}_t\bs{\nabla}_t \\
	\mb{a}_{t+1} &= \mb{c}+\mb{B}\mb{a}_t+\mb{B}\mb{P}_t\bs{\nabla}_t
\end{align}
whereas the backward recursions (\ref{eq:ksStandard1}), (\ref{eq:ksStandard2}) become:
\begin{align}
	\mb{r}_{t-1} &= \bs{\nabla}_t+\mb{L}_t'\mb{r}_t \\
	\hat{\bs{\alpha}}_t &= \mb{a}_t+\mb{P}_t\mb{r}_{t-1}
\end{align}
where, using (\ref{eq:app:info}) together with $\mb{K}_t=\mb{B}\mb{P}_t\mb{Z}'\mb{F}_t^{-1}$, the matrix $\mb{L}_t$ can be written as:
\begin{equation}
	\mb{L}_t=\mb{B}-\mb{K}_t\mb{Z}=\mb{B}-\mb{B}\mb{P}_t\mb{Z}'\mb{F}_t^{-1}\mb{Z}=\mb{B}-\mb{B}\mb{P}_t\bs{\mathcal{I}}_t
\end{equation}
Since the system matrices are constant, a steady state solution exists and $\mb{P}_t$ converges to the solution $\mb{\bar{P}}$ of the matrix Riccati equation (\ref{eq:riccati}) (\citealt{Harvey}, \citealt{DurbinKoopman}). Let us define $\mb{A}=\mb{B}\mb{\bar{P}}$, $\mb{\bar{F}}=\mb{Z}\mb{\bar{P}}\mb{Z}'+\mb{H}$ and $\bs{\mathcal{I}}=\mb{Z}'\mb{\bar{F}}^{-1}\mb{Z}$. Replacing $\mb{P}_t$ with $\mb{\bar{P}}=\mb{B}^{-1}\mb{A}$ and $\bs{\mathcal{I}}_t$ with $\bs{\mathcal{I}}$ in the four recursions above, we obtain:
\begin{align}
	\mb{a}_{t|t} &= \mb{a}_t+\mb{B}^{-1}\mb{A}\bs{\nabla}_t \\
	\mb{a}_{t+1} &= \mb{c}+\mb{B}\mb{a}_t+\mb{A}\bs{\nabla}_t
\end{align}
and
\begin{align}
	\mb{r}_{t-1} &= \bs{\nabla}_t+(\mb{B}-\mb{A}\bs{\mathcal{I}})'\mb{r}_t \\
	\hat{\bs{\alpha}}_t &= \mb{a}_t+\mb{B}^{-1}\mb{A}\mb{r}_{t-1}
\end{align}
which are Eq. (\ref{eq:kfNew1}), (\ref{eq:kfNew2}), (\ref{eq:ksNew1}), (\ref{eq:ksNew2}).
\begin{flushright}
$\square$
\end{flushright}

\section{The case with a scalar signal ($p=1$) and more latent components ($m>1$)}

\label{app:ident}
Let $y_t\in\mathbb{R}$, $\mb{a}_t\in\mathbb{R}^m$, $m>1$ and $\mb{Z}\in\mathbb{R}^{1\times m}$. Let us consider an observation density $p(y_t|\theta_t)$, where $\theta_t=\mb{Z}\mb{a}_t$ is a scalar signal. The score is given by:
\begin{equation}
	\bs{\nabla}_t=\frac{\partial\log p(y_t|\theta_t)}{\partial\mb{a}_t} = \mb{Z}'\nabla_t^{(\theta_t)}
\end{equation}
where $\nabla_t^{(\theta_t)}=\frac{\partial\log p(y_t|\theta_t)}{\partial\theta_t}$ is scalar. Similarly, the Fisher information matrix is given by:
\begin{equation}
	\bs{\mathcal{I}}_t = \Exp_{t-1}[\bs{\nabla}_t\bs{\nabla}_t']=\mb{Z}'\mb{Z}i_t^{(\theta_t)}
\end{equation}
where $i_t^{(\theta_t)}=\Exp_{t-1}[{\nabla_t^{(\theta_t)}}^2]$ is scalar.

Let us consider the case in which $\mb{a}_t$ evolves based on the score-driven scheme of Eq. (\ref{eq:sdsFgen}), written here in the notation of Section \ref{sub:genKf}:
\begin{equation}
\mb{a}_{t+1} = \mb{c} + \mb{B}\mb{a}_t + \mb{A}\bs{\nabla}_t
\end{equation}
where, since $\bs{\mathcal{I}}_t$ is singular, we have set $\mb{S}_t=\mb{I}$. Such choice for the normalization is equivalent to imposing a steady-state solution for the model.
The last term can be written as $\mb{a}\nabla_t^{(\theta_t)}$, where $\mb{a}=\mb{A}\mb{Z}'$. It follows that we can only identify the vector $\mb{a}$, but not the full matrix $\mb{A}$. Similarly, since $\mb{P}_t=\mb{B}^{-1}\mb{A}$, we can only identify the vector $\mb{P}_t\mb{Z}'$, but not the full matrix $\mb{P}_t$. This is not an issue when computing the two filters in Eq. \eqref{eq:kfGeneral2}, \eqref{eq:kfGeneral3}, as they only depend on $\mb{P}_t\mb{Z}'$. The same is true of the backward recursion \eqref{eq:ksGeneral1}, since $\mb{P}_t\bs{\mathcal{I}}_t=i_t^{(\theta_t)}(\mb{P}_t\mb{Z}')\mb{Z}$ is also determined by $\mb{P}_t\mb{Z}'$, so that the whole sequence $\mb{r}_t$ is identified. It is Eq. \eqref{eq:ksGeneral2} that is affected, since there $\mb{P}_t$ multiplies the generic vector $\mb{r}_{t-1}$, and we therefore obtain different smoothed estimates depending on the choice of $\mb{A}$ from the admissible set. The same argument can be generalized to the case $p>1$ and $m>p$.

\section{The $t$-GAS-SDS recursions}
\label{app:tgas}

This appendix reports the explicit form of the recursions of Section \ref{sub:sdsRec} for the $t$-GAS model of Section \ref{sec:tGAS}, which is the model employed in the empirical illustration of Section \ref{sec:empirics}. The score and the Fisher information matrix of the multivariate Student $t$ density are those derived by \cite{GAS2}; we restate them in the notation of Section \ref{sub:sdsRec} and combine them with the forward recursions \eqref{eq:sdsUgen}, \eqref{eq:sdsFgen} and the backward recursions \eqref{eq:sdsR_gen}, \eqref{eq:sdsS_gen}.

\paragraph{Notation.} We denote by $\bs{\mathcal{D}}_p$ the duplication matrix, defined by $\bs{\mathcal{D}}_p\text{vech}(\mb{X})=\text{vec}(\mb{X})$ for any symmetric $\mb{X}\in\mathbb{R}^{p\times p}$, by $\bs{\mathcal{B}}_p$ the elimination matrix, defined by $\bs{\mathcal{B}}_p\text{vec}(\mb{X})=\text{vech}(\mb{X})$, and by $\bs{\mathcal{C}}_p$ the commutation matrix, defined by $\bs{\mathcal{C}}_p\text{vec}(\mb{X})=\text{vec}(\mb{X}')$. For two matrices $\mb{X}$ and $\mb{Y}$ of the same order we write $\mb{X}\oplus\mb{Y}=(\mb{X}\otimes\mb{Y})+(\mb{Y}\otimes\mb{X})$, and we abbreviate $\mb{X}_{\otimes}=\mb{X}\otimes\mb{X}$; in particular $\mb{y}_{t\otimes}=\mb{y}_t\otimes\mb{y}_t$ and $\mb{V}_{t\otimes}^{-1}=\mb{V}_t^{-1}\otimes\mb{V}_t^{-1}$. Finally, $\mb{I}_p$ denotes the identity matrix of order $p$ and $\delta_{ij}$ the Kronecker delta.

\paragraph{Score and information matrix.} Let $\mb{V}_t=\mb{V}(\mb{f}_t)$ be a differentiable parameterization of the conditional covariance matrix in terms of the vector of time-varying parameters $\mb{f}_t\in\mathbb{R}^k$, and let
\begin{equation}
	\bs{\Psi}_t = \frac{\partial\,\text{vech}(\mb{V}_t)}{\partial\mb{f}_t'}\in\mathbb{R}^{\frac{p(p+1)}{2}\times k}
	\label{eq:tgasPsi}
\end{equation}
denote the associated Jacobian. For the observation density \eqref{eq:obs_tGAS}, \cite{GAS2} show that the score is given by
\begin{equation}
	\bs{\nabla}_t = \frac{\partial\log p(\mb{y}_t|\mb{V}_t,\nu)}{\partial\mb{f}_t} = \frac{1}{2}\bs{\Psi}_t'\bs{\mathcal{D}}_p'\mb{V}_{t\otimes}^{-1}\left[w_t\,\mb{y}_{t\otimes}-\text{vec}(\mb{V}_t)\right]
	\label{eq:tgasScore}
\end{equation}
where the scalar weight
\begin{equation}
	w_t = \frac{\nu+p}{\nu-2+\mb{y}_t'\mb{V}_t^{-1}\mb{y}_t}
	\label{eq:tgasWeights}
\end{equation}
plays the same role as the weight \eqref{eq:betatWeights} in the Beta-$t$-GARCH example: it downweights observations with a large Mahalanobis distance, and it is what makes the filter robust to outliers. The Fisher information matrix is
\begin{equation}
	\bs{\mathcal{I}}_t = \Exp_{t-1}[\bs{\nabla}_t\bs{\nabla}_t'] = \frac{1}{4}\bs{\Psi}_t'\bs{\mathcal{D}}_p'\mb{J}_{t\otimes}'\left[g\,\mb{G}-\text{vec}(\mb{I}_p)\text{vec}(\mb{I}_p)'\right]\mb{J}_{t\otimes}\bs{\mathcal{D}}_p\bs{\Psi}_t
	\label{eq:tgasInfo}
\end{equation}
where $\mb{J}_t$ is any square root of the precision matrix, $\mb{V}_t^{-1}=\mb{J}_t'\mb{J}_t$, the scalar $g$ is
\begin{equation}
	g = \frac{\nu+p}{\nu+2+p}
	\label{eq:tgasG0}
\end{equation}
and $\mb{G}\in\mathbb{R}^{p^2\times p^2}$ collects the fourth moments of a standard normal vector, with elements
\begin{equation}
	\mb{G}\left[(i-1)p+\ell,\;(j-1)p+m\right] = \delta_{ij}\delta_{\ell m}+\delta_{i\ell}\delta_{jm}+\delta_{im}\delta_{j\ell},\qquad i,j,\ell,m=1,\dots,p
	\label{eq:tgasG}
\end{equation}
Both \eqref{eq:tgasScore} and \eqref{eq:tgasInfo} depend on the chosen parameterization of $\mb{V}_t$ only through the Jacobian $\bs{\Psi}_t$, so that a change of parameterization only requires a new expression for $\bs{\Psi}_t$.

\paragraph{Hyperspherical parameterization.} We decompose the covariance matrix as $\mb{V}_t=\mb{D}_t\mb{R}_t\mb{D}_t$, where $\mb{D}_t$ is the diagonal matrix of conditional standard deviations and $\mb{R}_t$ is the conditional correlation matrix. Positive definiteness of $\mb{R}_t$ together with unit diagonal elements is imposed by writing $\mb{R}_t=\mb{X}_t'\mb{X}_t$, where $\mb{X}_t=\mb{X}(\bs{\phi}_t)$ is the upper triangular matrix with unit-norm columns
\begin{equation}
\mb{X}_t = \begin{pmatrix}
1 & c_{12t} & c_{13t} & \cdots & c_{1pt} \\
0 & s_{12t} & c_{23t}s_{13t} & \cdots & c_{2pt}s_{1pt} \\
0 & 0 & s_{23t}s_{13t} & \cdots & c_{3pt}s_{2pt}s_{1pt} \\
\vdots & \vdots & \vdots & \ddots & \vdots \\
0 & 0 & 0 & \cdots & \prod_{\ell=1}^{p-1}s_{\ell pt}
\end{pmatrix}
\label{eq:tgasX}
\end{equation}
with $c_{ijt}=\cos(\phi_{ijt})$ and $s_{ijt}=\sin(\phi_{ijt})$, and where $\bs{\phi}_t\in\mathbb{R}^{p(p-1)/2}$ collects the angles $\phi_{ijt}$, $i<j$. Combining this decomposition with a logarithmic specification of the variances, the vector of time-varying parameters is
\begin{equation}
	\mb{f}_t = \begin{pmatrix}\log\left(\text{diag}(\mb{D}_t^2)\right)\\ \bs{\phi}_t\end{pmatrix}\in\mathbb{R}^k,\qquad k = p+\frac{p(p-1)}{2}=\frac{p(p+1)}{2}
	\label{eq:tgasF}
\end{equation}
so that both blocks are unrestricted and the recursions can be run without imposing any constraint on $\mb{f}_t$. For this parameterization, \cite{GAS2} obtain
\begin{equation}
	\bs{\Psi}_t = \bs{\mathcal{B}}_p\left(\mb{I}_p\oplus\mb{D}_t\mb{R}_t\right)\mb{W}_{Dt}\mb{D}_t^2\bs{\mathcal{S}}_D + \bs{\mathcal{B}}_p\mb{D}_{t\otimes}\left[(\mb{I}_p\otimes\mb{X}_t')+(\mb{X}_t'\otimes\mb{I}_p)\bs{\mathcal{C}}_p\right]\mb{Z}_t\bs{\mathcal{S}}_{\phi}
	\label{eq:tgasPsiHyp}
\end{equation}
where $\bs{\mathcal{S}}_D$ and $\bs{\mathcal{S}}_{\phi}$ are the selection matrices extracting the two blocks of \eqref{eq:tgasF}, the matrix $\mb{W}_{Dt}$ is obtained from the $p^2\times p^2$ diagonal matrix with diagonal elements $\frac{1}{2}\text{vec}(\mb{D}_t^{-1})$ by deleting the columns containing only zeros, the factor $\mb{D}_t^2$ accounts for the logarithmic transformation of the variances, and
\begin{equation}
	\mb{Z}_t = \frac{\partial\,\text{vec}(\mb{X}_t)}{\partial\bs{\phi}_t'}\in\mathbb{R}^{p^2\times\frac{p(p-1)}{2}}
	\label{eq:tgasZ}
\end{equation}
collects the derivatives of \eqref{eq:tgasX}, which are available in closed form,
\begin{equation}
	\frac{\partial x_{ijt}}{\partial\phi_{\ell mt}} = \begin{cases}
	0 & \text{if } i>j,\ \ell\ge m,\ \ell\ge i,\ \text{or } j\ne m \\
	-x_{ijt}\tan(\phi_{ijt}) & \text{if } i=\ell \text{ and } i\ne j \\
	x_{ijt}/\tan(\phi_{\ell jt}) & \text{otherwise}
	\end{cases}
	\label{eq:tgasZel}
\end{equation}
$x_{ijt}$ denoting the $(i,j)$ element of $\mb{X}_t$.

\paragraph{Filtering and smoothing recursions.} Following \cite{GAS2}, we scale the score by the inverse information matrix, $\mb{S}_t=\bs{\mathcal{I}}_t^{-1}$, so that $\mb{s}_t=\bs{\mathcal{I}}_t^{-1}\bs{\nabla}_t$. The forward recursions \eqref{eq:sdsUgen}, \eqref{eq:sdsFgen} then read
\begin{align}
	\mb{f}_{t|t} &= \mb{f}_t + \mb{B}^{-1}\mb{A}\bs{\mathcal{I}}_t^{-1}\bs{\nabla}_t \label{eq:tgasU}\\
	\mb{f}_{t+1} &= \bs{\omega} + \mb{A}\bs{\mathcal{I}}_t^{-1}\bs{\nabla}_t + \mb{B}\mb{f}_t \label{eq:tgasFilt}
\end{align}
$t=1,\dots,n$, with $\bs{\nabla}_t$ and $\bs{\mathcal{I}}_t$ given by \eqref{eq:tgasScore} and \eqref{eq:tgasInfo}, and $\bs{\Psi}_t$ by \eqref{eq:tgasPsiHyp}. The predictive filter \eqref{eq:tgasFilt} is the $t$-GAS model of \cite{GAS2}. Since $\mb{S}_t=\bs{\mathcal{I}}_t^{-1}$, the information matrix cancels from the backward recursion \eqref{eq:sdsR_gen}, which reduces to
\begin{align}
	\mb{r}_{t-1} &= \bs{\mathcal{I}}_t^{-1}\bs{\nabla}_t + (\mb{B}-\mb{A})'\mb{r}_t \label{eq:tgasR}\\
	\hat{\mb{f}}_t &= \mb{f}_t + \mb{B}^{-1}\mb{A}\mb{r}_{t-1} \label{eq:tgasS}
\end{align}
$t=n,\dots,1$, with $\mb{r}_n=\mb{0}$. As is customary in multivariate score-driven models, $\mb{A}$ and $\mb{B}$ are restricted to be diagonal, so that the number of static parameters grows linearly rather than quadratically in $k$.

Three remarks are in order. First, the Gaussian limit is recovered as $\nu^{-1}\rightarrow0$. In that case $w_t\rightarrow1$ and $g\rightarrow1$, the score \eqref{eq:tgasScore} reduces to the score of the Gaussian density and \eqref{eq:tgasInfo} to the corresponding information matrix, so that \eqref{eq:tgasU}--\eqref{eq:tgasS} become the filtering and smoothing recursions of the Gaussian GAS covariance model. This is the multivariate counterpart of the limit discussed after Eq. \eqref{eq:betatWeights}.

Second, the choice of the parameterization is not innocuous for the smoother. Under the alternative decomposition of \cite{GAS2}, in which the correlations are driven by $\text{vech}(\mb{Q}_t)$ as in the DCC model of \cite{EngleDCC}, the number of time-varying parameters is $k=p+p(p+1)/2$ and thus exceeds the dimension $p(p+1)/2$ of the signal $\text{vech}(\mb{V}_t)$ entering the observation density. The Jacobian $\bs{\Psi}_t$ does not have full column rank, the information matrix \eqref{eq:tgasInfo} is singular, and \cite{GAS2} scale the score through its Moore--Penrose pseudoinverse. This is exactly the situation discussed in Remark \ref{rem:ident}. The two filters and the sequence $\mb{r}_t$ are unaffected, because they depend on $\mb{A}$ only through identified combinations, but the smoothed estimate \eqref{eq:tgasS} is not uniquely defined, since there $\mb{A}$ multiplies the generic vector $\mb{r}_{t-1}$. Under the hyperspherical parameterization the $p(p-1)/2$ angles are instead in one-to-one correspondence with the $p(p-1)/2$ free correlations, $k$ equals the dimension of the signal, $\bs{\mathcal{I}}_t$ is nonsingular and the difficulty does not arise. This is a further reason, in addition to the quality of the estimates, for adopting it in the empirical illustration of Section \ref{sec:empirics}.

Third, the backward pass remains free in the sense of Appendix \ref{app:cost}. The expensive object in this model is the information matrix \eqref{eq:tgasInfo}, whose construction involves the $p^2\times p^2$ matrix $\mb{G}$; but $\bs{\mathcal{I}}_t$ is already required by the forward recursion in order to scale the score, and is stored when going forward. The recursions \eqref{eq:tgasR}, \eqref{eq:tgasS} evaluate no density and recycle $\bs{\nabla}_t$ and $\bs{\mathcal{I}}_t$, so that smoothing adds only matrix-vector products of order $k^2$ per period. This is what makes the SDS applicable at the cross-sectional dimensions considered in Section \ref{sec:empirics}, where simulation-based smoothing would be problematic.

\end{document}